\RequirePackage{fix-cm}
\documentclass[twocolumn,epjc3]{svjour3}
\smartqed  
\RequirePackage{graphicx}
\usepackage{epsfig}
\usepackage{bm} 
\usepackage{amssymb} 
\usepackage[usenames]{color}
\usepackage{bbding}
\usepackage{slashed}
\usepackage{array}
\usepackage{cite}  
\newcommand{\Tr}{\mbox{Tr}}
\newcommand{\vphi}{\varphi}
\newcommand{\Lagr}{{\cal L}}
\newcommand{\cO}{{\cal O}}

\newcommand{\MMC}{\mbox{\boldmath $M^2_\mathrm{C}$}}
\newcommand{\MMN}{\mbox{\boldmath $M^2_\mathrm{N}$}}

\newcommand{\omegaPP}{\bar{\omega}^{\perp}}
\newcommand{\omegaPL}{\bar{\omega}^{\parallel}}
\newcommand{\bgs}{B\rightarrow X_s\gamma}

\newcommand{\gap}{\mbox{\hspace{0.5cm}}}
\newcommand{\pard}{\partial}
\newcommand{\BB}{\mbox{\boldmath$B$}}
\newcommand{\BW}{\mbox{\boldmath$W$}}
\newcommand{\BV}{\mbox{\boldmath$V$}}
\newcommand{\cL}{{\cal L}}
\newcommand{\cM}{{\cal M}}

\newcommand{\Comm}[2]{\mbox{$[$} #1, #2\mbox{$]$}}

\newcommand{\MatrixTwo}[4]
{\left( \begin{array}{cc}
#1 & #2 \\
#3 & #4
\end{array}\right)}

\newcommand{\MatrixThree}[9]
{\left( \begin{array}{ccc}
#1 & #2 & #3 \\
#4 & #5 & #6 \\
#7 & #8 & #9
\end{array}\right)}
\newcommand{\unit}[1]{\;\mathrm{#1}}
\journalname{Eur. Phys. J. C}
\begin{document}

\title{The limits on the strong Higgs sector parameters \\
       in the presence of new vector resonances
}


\author{Mikul\'{a}\v{s} Gintner\thanksref{e1,addr1,addr2}
        \and
        Josef Jur\'{a}\v{n}\thanksref{e2,addr2,addr3} 
}

\thankstext{e1}{e-mail: gintner@fyzika.uniza.sk}
\thankstext{e2}{e-mail: josef.juran@utef.cvut.cz}


\institute{Physics Department, University of \v{Z}ilina,
Univerzitn\'{a} 1, 010 26 \v{Z}ilina, Slovakia \label{addr1}
           \and
           Institute of Experimental and Applied Physics,
Czech Technical University in Prague, Horsk\'{a} 3a/22, 128 00
Prague, Czech Republic \label{addr2}
           \and
           Institute of Physics, Silesian University in Opava,
Bezru\v{c}ovo n\'{a}m. 13, 746 01 Opava, Czech Republic
\label{addr3} }

\date{Received: date / Accepted: date}

\maketitle

\begin{abstract}
In this paper, we investigate how the LHC data limit the Higgs-related
couplings in the effective description of a strongly interacting extension of
the Standard model. The Higgs boson is introduced as a scalar composite state and
it is followed in the mass hierarchy by an $SU(2)$ triplet of vector composites.
The limits are calculated from the constraints obtained in the recent ATLAS+CMS combined analysis of
the data from 2011 and 2012. We find that the data prefer the scenario
where the Higgs couplings to the electroweak gauge bosons differ from its
couplings to the vector triplet. 
We also investigate the unitarity limits of the studied effective model
for the experimentally preferred values of the Higgs couplings. We find from
the $\pi\pi\rightarrow\pi\pi$ scattering amplitudes that for the vector resonance masses
between one and two TeV significant portions of the experimentally allowed regions
are well below the unitarity limit.
We also evaluate how the existing ATLAS and CMS Run-2 data restrict our model with the upper bounds
on the resonance production cross section times its branching ratio for various decay
channels.
The masses
in the range $1\unit{TeV}\leq M_\rho\leq 2\unit{TeV}$ are not excluded
in parts or even full parameter space of our theory.
\end{abstract}

\section{Introduction}
\label{sec:intro}
Even though the LHC experiments ATLAS and CMS achieved a spectacular 
success by discovering the 125 GeV Higgs boson~\cite{125GeVBosonDiscoveryATLAS,125GeVBosonDiscoveryCMS} 
it was more the beginning
rather than the end of the struggle to uncover the character of physics beyond 
the Standard model (SM). To this moment, it has not even been settled down whether 
new physics takes the form of weakly coupled supersymmetry or strongly coupled composites.

In the strongly coupled scenario, the observed lightness of the Higgs boson with respect
to the expected size of the compositeness scale (naively, $\Lambda=4\pi v\simeq 3\unit{TeV}$) 
could be explained if the Higgs boson were a (pseudo)Nambu--Goldstone boson after 
a breakdown of suitable extended global 
symmetry~\cite{Barbieri_etal07,CsakiFalkowskiWeiler08,Contino10,Contino_etal11,PomarolRiva12,
PappadopuloThammTorre13,Montull_etall13,PanicoWulzer16}.
Another widely studied possibility is to embed the composite Higgs in an effective 
$SU(2)$ doublet~\cite{BardeenHillLindner90,Giudice_etal07,
Foadi_etal07,RyttovSannino08,Sannino09,Zerwekh10,
HernandezTorre10,BurdmanHaluch11,HapolaSannino11,Hernandez_etal12,Foadi_etal13,Contino_etal13,
Castillo-Felisola_etal13,
HernandezDibZerwekh14,Pappadopulo_etal14,Belyaev_etal14} where its lightness would be
guaranteed by the theory's particularities.

If the Higgs is generated as a composite state by new strong interactions
the extension of the SM can be effectively described by higher dimensional
operators that do not decouple in the low-energy limit. Presumably, they would modify
the SM couplings of the Higgs boson with the heavy SM fields, such as the electroweak (EW)
gauge bosons and/or
the third quark generation. However, while the light SM Higgs boson can guarantee
unitarity of the SM to virtually arbitrary high energies, this is not true anymore
if the Higgs couplings become anomalous~\cite{LeeQuiggThacker77,Chanowitz85}. 
Nevertheless, the least one could require from the successful
effective description of the composite state phenomenology is that it will not break down
at energy below the compositeness scale. Meeting this expectation might be 
assisted with by the presence of additional new composite states which naturally occur in 
strongly interacting theories, anyway.
Any further progress in dealing with these questions largely depends on an experimental 
input. Therefore, it is interesting to find what the most recent measurements of the
Higgs boson couplings imply for various effective descriptions of strongly interacting 
extensions of the SM.

In this paper, we calculate the LHC limits on the effective theory describing possible early signs
of strongly interacting physics beyond the SM. The effective description we work with 
is a rather simplistic view of what might be observed at the LHC beyond the 125 GeV Higgs boson. 
We work with the vision
where the Higgs boson is a scalar composite state followed in the mass hierarchy by
a vector composite $SU(2)$ triplet state. 
Our approach is
closely related to the formalism used in~\cite{CompositeHiggsSketch}.

In particular, the Higgs sector of the effective Lagrangian under consideration is based on 
the non-linear sigma model with the 125-GeV $SU(2)_{L+R}$ scalar singlet complementing 
the non-linear triplet of the Nambu--Goldstone bosons. The new vector resonances are 
explicitly present in the form of an $SU(2)_{L+R}$ triplet. The vector triplet is introduced as a
gauge field via the hidden local symmetry approach~\cite{HLS}.
Consequently, it mixes with the EW gauge bosons.
It results in the appearance of the mixing-generated (indirect) couplings of the vector triplet
with all SM fermions.
This setup fits the situation when the global
$SU(2)_L\times SU(2)_R$ symmetry is broken down to $SU(2)_{L+R}$.
The gauge sector of this effective description is equivalent to the gauge sector of
highly deconstructed Higgsless model with only three sites~\cite{3siteHiggslessModel}.

The above-mentioned effective scenario has also been a basis for the effective Lagrangian
we introduced and investigated in~\cite{tBESSprd11,tBESSepjc13}\footnote{
We call it the tBESS model
to stress its connection to the BESS model~\cite{BESS}. 
The name modifying ``t'' suggests a special
standing of the top-quark related doublet in the model.}.
Therein, the vector triplet couples directly to the third quark doublet
only and to none of the other SM fermions. 
In addition, even the interactions to the right quarks
are not necessarily universal. 
Similar interaction patterns can be found in various recent extensions of the SM,
including the partial compositeness and extra-dimensional scenarios.
As was shown in~\cite{tBESSprd11, tBESSepjc13} 
this arrangement helps relax the tight restrictions
placed by the electroweak precision data on the vector triplet coupling
to the light SM fermions, the bottom quark included. While the study in this paper 
follows our tBESS model (and, in the relevant parts, matches the model considered
in~\cite{CompositeHiggsSketch}), its conclusions will also be partly, or fully, applicable to 
a wider class of effective models, e.g., with different fermionic sectors.

The question we address in this paper concerns a possible structure of the interactions
between the new vector triplet and the Higgs boson. Under the considered symmetry,
the Higgs boson coupling to the new vector triplet can differ from the Higgs couplings
to the EW gauge bosons. This splitting might appear as an unwelcome complication.
Nevertheless, as we will demonstrate in the paper, the ATLAS and CMS data
support it. In particular, the $H\rightarrow\gamma\gamma$ constraint is 
the key component of the data that makes it difficult for the no-splitting scenario 
to satisfy the LHC measurements.

Using the results of~\cite{CompositeHiggsSketch,tBESSprd11,tBESSepjc13} we also analyze
the tree-level unitarity limits of our model resulting from the scattering of 
the longitudinal EW gauge bosons for the allowed values of the Higgs couplings
obtained in this work. When authors of~\cite{CompositeHiggsSketch} addressed unitarity questions
the discovery of the Higgs boson was not confirmed yet. Thus, while they used the correct 
mass of the Higgs boson in their analysis, they were lacking any of the experimental input on
the Higgs interactions available to us today. This was compensated for by the usage of
the sum rules. However, one of the used sum rules imposes the no-spitting condition.
The results of our analysis compel us to abandon this assumption and calculate the unitarity
limits under different conditions.

This paper is organized as follows.
In Section~\ref{sec:efflagr}, we introduce the necessary components of the effective
Lagrangian and work out the consequences of the Higgs-to-gauge-boson coupling splitting.
Section~\ref{sec:LimitsHiggsRelatedParams} is devoted to the calculations of the constraints
for the Higgs-to-gauge boson couplings. Particularly, 
in Subsection~\ref{subsec:relevant-measurements},
we set up the framework for the constraint calculations. In Subsection~\ref{subsec:aV-equals-arho},
we demonstrate the tension between the $H\rightarrow\gamma\gamma$ and other LHC measurement 
constraints for the model parameters when the universality of the Higgs-to-gauge-boson couplings
is assumed. In Subsection~\ref{subsec:CombinedAnalysis}, we calculate the best fits
and constraints for the Higgs-to-gauge-boson parameters when the Higgs coupling to the new
vector triplet can differ from the Higgs coupling to the EW gauge bosons.
In Subsection~\ref{subsec:UnitarityLimits} we investigate the unitarity limits for
our model.
Finally, in Section~\ref{sec:LimitsVectorMassAndXS}, we investigate the applicability of 
the existing LHC limits on the masses of new vector resonances to our model. 
We compare the predictions of our model for the production cross section of the vector
resonance times its branching ratio for various decay channels with the existing 
experimental upper bounds obtained by the ATLAS and CMS Collaborations.
Section~\ref{sec:conclusions} presents the conclusions of the paper.


\section{The effective Lagrangian}
\label{sec:efflagr}
The effective Lagrangian
is built to respect the global $SU(2)_L\times SU(2)_R\times U(1)_{B-L}\times
SU(2)_{HLS}$ symmetry of which the $SU(2)_L\times
U(1)_Y\times SU(2)_{HLS}$ subgroup is also a local symmetry.
The $SU(2)_{HLS}$ symmetry is an auxiliary gauge symmetry
invoked to accommodate the $SU(2)$
triplet of vector resonances. Beside the scalar singlet $h(x)$ and
the vector triplet $\vec{V}_\mu=(V_\mu^1,V_\mu^2,V_\mu^3)$, 
the effective Lagrangian is built out of the SM fields only.

The Lagrangian can be split in three terms\footnote{
While the full formulation of the model can be found in~\cite{tBESSepjc13},
the definitions of basic quantities used in Eqs.~(\ref{eq:LagGB})
through (\ref{eq:LagFermScalar}) are, for the reader's convenience, summarized
in~\ref{app:definitions}.}
\begin{equation}\label{eq:LagTBESS}
  \Lagr = \Lagr_\mathrm{GB} + \Lagr_\mathrm{ESB} + \Lagr_\mathrm{ferm},
\end{equation}
where $\cL_\mathrm{GB}$ describes the gauge-boson sector including
the $SU(2)_\mathrm{HLS}$ triplet, 
\begin{eqnarray}\label{eq:LagGB}
   \Lagr_\mathrm{GB} &=& \frac{1}{2g^2}\mathrm{Tr}(\BW_{\mu\nu}\BW^{\mu\nu}) 
   +\frac{1}{2g^{\prime 2}}\mathrm{Tr}(\BB_{\mu\nu}\BB^{\mu\nu}) 
   \nonumber\\ &&
   +\frac{2}{g^{\prime\prime 2}}\mathrm{Tr}(\BV_{\mu\nu}\BV^{\mu\nu}),
\end{eqnarray}
$\cL_\mathrm{ESB}$ is the scalar sector
responsible for spontaneous breaking of the electroweak and
hidden local symmetries, and $\cL_\mathrm{ferm}$ is the fermion Lagrangian
of the model.

Let us express $\Lagr_\mathrm{ESB}$ as a sum of two terms, 
$\Lagr_\mathrm{ESB}= \Lagr_h + \Lagr_\mathrm{hV}$, where 
\begin{equation}\label{eq:Lagh}
   \Lagr_h = \frac{1}{2}\pard_\mu h \pard^\mu h - \frac{1}{2}M_h^2 h^2 - c_h\;\frac{M_h^2}{2v}h^3 
                                  - c_h'\;\frac{M_h^2}{8v^2}h^4
\end{equation}
contains the kinetic term and the self-interactions of the Higgs boson
with the mass $M_h=125\unit{GeV}$ and free parameters $c_h$ and $c_h'$. Furhter,
\begin{eqnarray}\label{eq:Lag2}
   \Lagr_\mathrm{hV} &=& -v^2\left[\Tr(\omegaPP)^2(1+2a_V\frac{h}{v}+a_V'\frac{h^2}{v^2}+\ldots)\right.
   \nonumber\\ && \phantom{-v^2[}
    \left.    +\alpha\Tr(\omegaPL)^2(1+2a_\rho\frac{h}{v}+a_\rho'\frac{h^2}{v^2}+\ldots)\right]
\end{eqnarray}
is responsible for the masses of all gauge bosons including the new vector triplet, and
describes their interactions with the Higgs boson. The interactions are parameterized by
the free parameters $a_V, a_\rho, a_V', a_\rho',\ldots$.
Below, only the interaction terms of (\ref{eq:Lag2}) that are at most linear in $h$ 
will be considered\footnote{Although the quadratic terms would be needed if one
wished to maintain the possibility to eliminate the linear growth in $s$ from the scattering
amplitude $\pi\pi\rightarrow hh$ and thus improve the unitarity limit for the model.}.  
Terms with higher powers of $h$, 
which are not important for higgs phenomenology at the LHC, can be restored at any time without
affecting our conclusions.

The fermion sector $\Lagr_\mathrm{ferm}$ in its minimal formulation contains 
the fermionic kinetic terms and the fermion interactions with the EW gauge bosons
as well as terms responsible for the couplings of the SM fermions to the Higgs boson.
While the EW part is kept identical to the SM one, the interactions of the Higgs
boson with the fermions can assume non-SM values. Their parameterization is based
on the following interaction Lagrangian:
\begin{equation}\label{eq:LagFermScalar}
    \Lagr_\mathrm{ferm}^\mathrm{scalar} = -\sum_{k=1}^6 \bar{\psi}_L^k U M_f^k (1+c_f^k\frac{h}{v}
               +c_f^{\prime k}\frac{h^2}{v^2}+\ldots)\psi_R^k + \mathrm{h.c.},
\end{equation}
where $M_f^k$ is a $2\times 2$ diagonal matrix with the masses of
the upper and bottom $k^\mathrm{th}$ fermion doublet components on its diagonal,
and $U=\xi(\vec{\pi})\cdot\xi(\vec{\pi})=\exp(2i\vec{\pi}\vec{\tau}/v)$.
Note that when $c_k=1, \forall k$, and the rest of $c$'s
are zeros the scalar resonance interactions with fermions imitate
the corresponding interactions of the SM Higgs boson. Again,
only the interaction terms that are at most linear in $h$ 
will be considered below.

Possible direct interactions of the fermions with the vector triplet do not play
a role in the calculation of the limits investigated in this paper. Thus, this part 
of the effective Lagrangian is left unspecified. Nevertheless, we would 
like to mention the example of the setup of this sector where
only the third quark generation couples directly to the vector triplet and the interaction
of the right top quark is disentangled from the interaction of 
the right bottom quark. We suggested and analyzed the effective model with
this kind of the fermion sector in~\cite{tBESSprd11,tBESSepjc13}.
For the sake of completeness, we would like to point out that in this model the vector resonances 
do interact with the SM fermions, including the light ones, even if there are no 
direct interactions introduced. This is due to the mixing between the EW gauge bosons
and the vector resonance triplet. Of course, the mixing-induced couplings are suppressed
by the transformation matrix elements; they are proportional to $1/g''$.

The mixing of the gauge fields occurs in the process of diagonalization of the gauge-boson
mass matrix. After gauging out all six Goldstone bosons the Lagrangian $\Lagr_\mathrm{hV}$
reads
\begin{equation}
   \Lagr_\mathrm{hV} = \cM(\alpha) + \frac{2a_V}{v}\cM(\alpha r) h,
\end{equation}
where $r=a_\rho/a_V$ 
and where
\begin{eqnarray}
   \cM(\alpha) &\equiv&  \frac{1}{2} (X_{\mu}^-)^\dagger\cdot \MMC(\alpha)\cdot X_{\mu}^- 
                  +  \frac{1}{2} (X_{\mu}^+)^\dagger\cdot \MMC(\alpha)\cdot X_{\mu}^+
   \nonumber\\ &&
                   +  \frac{1}{2} (X_{\mu}^0)^\dagger\cdot \MMN(\alpha) \cdot X_{\mu}^0
\end{eqnarray}
is the gauge-boson mass term. Further, $X^\pm = (W^\pm_f, V^\pm)^T$, $X^0 = (W^3, B, V^3)^T$,
and $\MMC(\alpha)$ and $\MMN(\alpha)$ are the squared-mass matrices of the charged and 
neutral gauge bosons, respectively,
\begin{eqnarray}
   \MMC &=& \frac{v^2}{4}\MatrixTwo{(1+\alpha)g^2}{-\alpha gg''}
                                         {-\alpha gg''}{\alpha g^{\prime\prime 2}}, \\              
   \MMN &=& \frac{v^2}{4}\MatrixThree{(1+\alpha)g^2}{-(1-\alpha)gg'}{-\alpha gg''}
                                           {-(1-\alpha)gg'}{(1+\alpha)g^{\prime 2}}{-\alpha g'g''}
                                           {-\alpha gg''}{-\alpha g'g''}{\alpha g^{\prime\prime 2}}.
\end{eqnarray}
The diagonalization process results in the transformation
of the gauge-boson basis, from the mass one to the flavor one, $\{Y\}\rightarrow \{X\}$:
\begin{equation}
   X_{\mu}^\pm = O^{(C)}\cdot Y_{\mu}^\pm,\hspace{0.5cm}  X_{\mu}^0 = O^{(N)}\cdot Y_{\mu}^0,
\end{equation}
where $Y^\pm = (W^\pm_m,\rho^\pm)^T$, $Y^0 = (A,Z,\rho^0)^T$. Note that we use $V_\mu^a$
to denote the vector resonance components in the flavor basis and $\rho_\mu^a$ for 
the vector resonance components in the mass basis\footnote{We also used 
the subscripts $m$ and $f$
to distinguish the components of $W^\pm$ fields in the two bases. We do not use the subscripts
if the choice of the basis is obvious from the context.}.

In the limit $M_{W^\pm}, M_Z \ll M_{\rho^0}$, equivalent to the condition 
$g, G\ll \sqrt{\alpha}g''$, the mixing matrices read
\begin{eqnarray}
   O^{(C)}&=&\MatrixTwo{1}{-g/g''}{g/g''}{1}, \\
   O^{(N)}&=&\MatrixThree{g'/G}{g/G}{-g/g''}
                       {g/G}{-g'/G}{-g'/g''}
                       {2\frac{gg'}{Gg''}}{\frac{g^2-g^{\prime 2}}{Gg''}}{1}.
\end{eqnarray}
In the same limit, the next-to-leading order approximations of the gauge-boson masses read
\begin{eqnarray}
   M_{W^\pm} &=& \frac{vg}{2}\left(1-\frac{g^2}{2g^{\prime\prime 2}}\right), \\
   M_{\rho^\pm}&=&\frac{\sqrt{\alpha}vg''}{2}\left(1+\frac{g^2}{2g^{\prime\prime 2}}\right),
\end{eqnarray}
and
\begin{eqnarray}
   M_Z &=& \frac{vG}{2}\left[1-\frac{(g^2-g^{\prime 2})^2}{2g^{\prime\prime 2}G^2}\right], \\
   M_{\rho^0} &=& \frac{\sqrt{\alpha}vg''}{2}\left(1+\frac{G^2}{2g^{\prime\prime 2}}\right).
\end{eqnarray}
The leading order approximation for the partial width of the $\rho$ decay to the EW gauge bosons is
\begin{equation}\label{eq:DecayWidth}
   \Gamma(\rho^0\rightarrow W^+W^-) = \Gamma(\rho^\pm\rightarrow W^\pm Z) 
   = \frac{M_\rho^5}{48\pi v^4 g^{\prime\prime 2}}.
\end{equation}

The interactions of the gauge bosons with the Higgs can be read off from $\Lagr_\mathrm{hV}$
in the mass basis
\begin{eqnarray}\label{eq:HcplngLinear}
   \frac{2a_V}{v}\cM(\alpha r)h &=&  \frac{2h}{v}\left[
   \frac{1}{2}c_Z M_Z^2Z_\mu Z^\mu+c_W M_W^2W^+_\mu W^{-\mu} \right.
   \nonumber \\
   && \left.\phantom{\frac{2h}{v}\left[ \right. }
   +\frac{1}{2}c_{\rho^0}M_{\rho^0}^2\rho^0_\mu \rho^{0\mu}
   +c_{\rho^\pm}M_{\rho^{\pm}}^2\rho^+_\mu \rho^{-\mu} \right.
   \nonumber\\
   && \left.\phantom{\frac{2h}{v}\left[ \right. }
   +c_{W\rho^\pm}M_W M_{\rho^\pm}(W^+_\mu\rho^{-\mu}+\mathrm{h.c.})\right.
   \nonumber\\
   && \left.\phantom{\frac{2h}{v}\left[ \right. }
   +c_{Z\rho^0}M_Z M_{\rho^0}Z_\mu\rho^{0\mu} \right]
\end{eqnarray}
with the anomalousness factors\footnote{
If $Y=Y'$ then a single letter subscript will be used, e.g.\
$c_{YY}\rightarrow c_Y$.} 
$c_{YY'}$. If the flavor basis splitting factor $r$
equals to 1 ($a_\rho=a_V$) then the mass basis couplings $c_{YY'}$ follow a
simple pattern
\begin{equation}
   c_Z = c_W = c_{\rho^0} = c_{\rho^\pm} = a_V,\gap 
   c_{Z\rho^0} = c_{W\rho^\pm} =0.
\end{equation}

For a more general situation, $a_\rho\neq a_V$, the relations of $c_{YY'}$'s to $a_V$ 
and $a_\rho$ become more intricate.
Let us introduce the mass basis splitting factors $\zeta_{YY'}$ such that
\begin{equation}\label{eq:cecka}
   c_{YY'} =  a_V\;\zeta_{YY'}.
\end{equation} 
Then
\begin{eqnarray}
 \zeta_{YY'}(\alpha,r) &=& \frac{
   O_{Y_1Y}(\alpha) \cdot [\mbox{{\boldmath $M^2_\mathrm{C,N}$}}(\alpha r)]_{Y_1Y_2}
   \cdot O_{Y_2Y'}(\alpha)      }
 {M_Y(\alpha) M_{Y'}(\alpha)} \;
\end{eqnarray}
The $\zeta_{YY'}$ factors for the individual gauge bosons are summarized 
in Table~\ref{tab:SplittingFactors}.
\begin{table*}[t]
\centering
\caption{\label{tab:SplittingFactors}
         The mass basis splitting factors $\zeta_{YY'}(\alpha,r)$ for the individual gauge bosons
         $Y$, $Y'\in \{W^\pm,\rho^\pm,A,Z,\rho^0\}$.
        }
{\renewcommand{\arraystretch}{1.6}
\begin{tabular}{c|ccccc}       
   \hline
   $\zeta_{YY'}(\alpha,r)$ & $\rho^0$ & $Z$ & $A$ & $\rho^\pm$ & $W^\pm$ \\
   \hline
   $W^\pm$ & 0 & 0 & 0 & $g^2\sqrt{\frac{\alpha}{D_C}}(r-1)$ & $\frac{1+r +(1-r)k_C}{2}$ \\
   $\rho^\pm$ & 0 & 0 & 0 & $\frac{1+r -(1-r)k_C}{2}$ & \multicolumn{1}{c}{} \\
   $A$ & 0 & 0 & 0 &  \multicolumn{2}{c}{} \\
   $Z$ & $(g^2-g^{\prime 2})\sqrt{\frac{\alpha}{D_N}}(r-1)$ & $\frac{1+r +(1-r)k_N}{2}$ & \multicolumn{3}{c}{} \\
   $\rho^0$ & $\frac{1+r -(1-r)k_N}{2}$ & \multicolumn{4}{c}{} \\
\end{tabular}}
\end{table*}
There, we have introduced auxiliary variables
\begin{eqnarray}
 \sqrt{D_N} &=& \frac{4}{v^2}\left(M_{\rho^0}^2-M_Z^2\right), \\
 \sqrt{D_C} &=& \frac{4}{v^2}\left(M_{\rho^\pm}^2-M_W^2\right),
\end{eqnarray}
and
\begin{eqnarray}
   k_N &=& [1-4\alpha (g^2-g^{\prime 2})^2/D_N]^{1/2}, \\
   k_C &=& (1-4\alpha g^4/D_C)^{1/2}.
\end{eqnarray}
Note that $k_{N,C}=1-\cO(x^4)$ where 
$x=g/(\sqrt{\alpha}g'')\approx M_W/M_\rho$.

\begin{figure}
\centering
\includegraphics[scale=0.85]{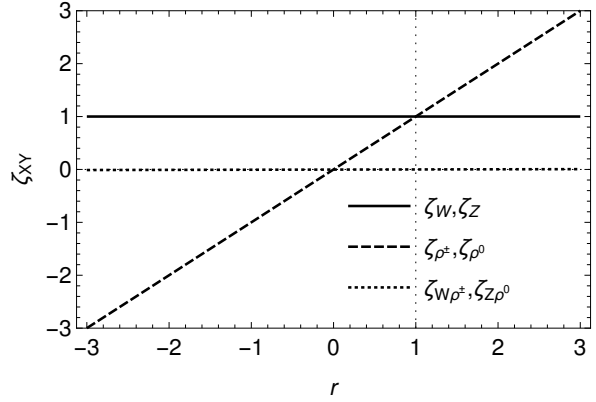}
\caption{\label{fig:SplittingImpactOnAnomCplngs}
         The mass basis splitting factors $\zeta_{XY}$ as functions of
         the splitting $r=a_\rho/a_V$. 
         The graphs are plotted for $M_{\rho^0} = 1.5\unit{TeV}$ and $g''=15$.
         }
\end{figure}

Fig.~\ref{fig:SplittingImpactOnAnomCplngs} helps to understand how the flavor basis splitting,
$r=a_\rho/a_V$, translates into the anomalous factors $c_{YY'}$. 
The essential role in this issue is played by the $\zeta$ factors.
Therefore, the graphs in the figure depict the dependences of $\zeta$'s on $r$.
There, the plots of $\zeta_{W,Z}$ are almost perfect horizontal 
lines at 1 which complies with $\zeta_{W,Z}=1-(1-r)\cO(x^4)$.
On the other hand, $\zeta_{\rho^\pm,\rho^0}=r+(1-r)\cO(x^4)$ suggests 
the straight line of the $45$ degree slope for $\zeta_{\rho^\pm,\rho^0}(r)$.
Finally, the $\zeta$ factors of the mixed interaction terms are negligible when compared to 
the other $\zeta$'s, which is in agreement with the finding that 
$\zeta_{W\rho^\pm,Z\rho^0} = (1-r)\cO(x^2)$. 

\begin{figure}
\centering
\includegraphics[scale=0.85]{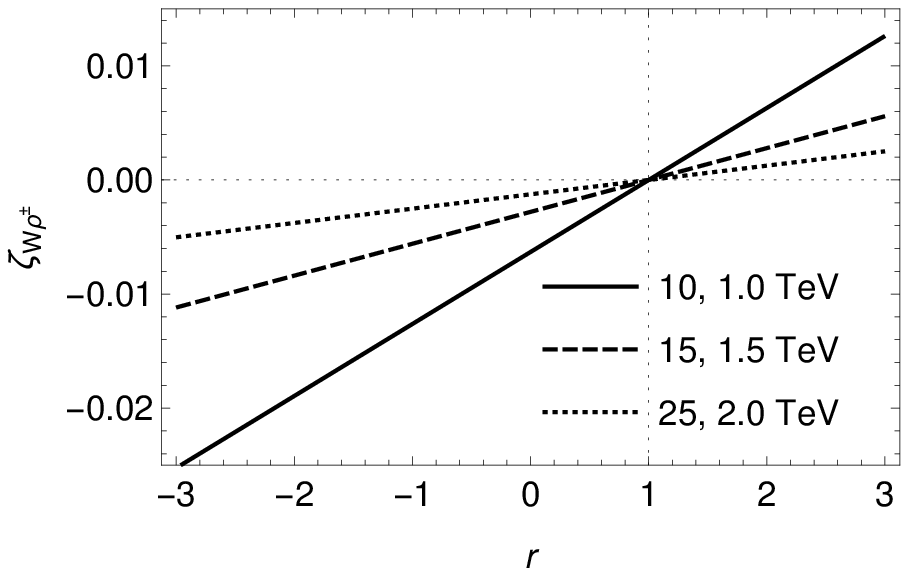}
\includegraphics[scale=0.85]{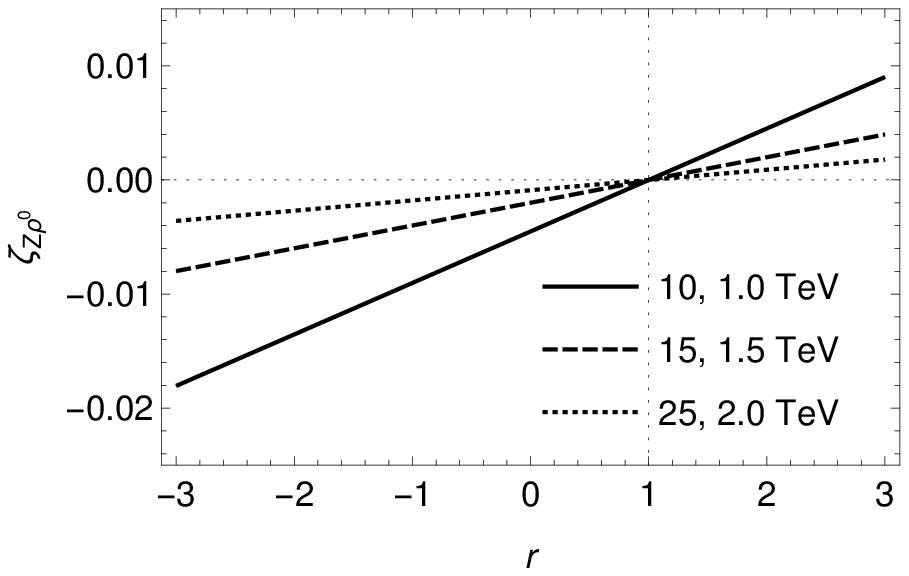}
\caption{\label{fig:NondiagZetas}
         The non-diagonal splitting factors $\zeta_{W\rho^\pm}$ (left panel) and 
         $\zeta_{Z\rho^0}$ (right panel)
         as functions of
         the splitting $r=a_\rho/a_V$.
         The solid lines correspond to $(g'',M_\rho)=(10,1\unit{TeV})$,
         the dashed lines to $(15,1.5\unit{TeV})$, and 
         the dotted lines to $(25,2\unit{TeV})$.
         }
\end{figure}

While all $\zeta$'s depend on $g''$ and $M_\rho$ these dependences are very weak.
When changing $(g'',M_\rho)$ from $(10,1\unit{TeV})$ to $(25,2\unit{TeV})$,
$\zeta_W$ and $\zeta_Z$ vary no more than by about $10^{-4}$ for $-3\leq r\leq 3$. 
The same conclusion applies to
$\zeta_{\rho^\pm}$ and $\zeta_{\rho^0}$. The actual size of $\zeta_{W\rho^\pm}$
and $\zeta_{Z\rho^0}$ as well as their dependence on $g''$ and $M_\rho$ can be seen
in Fig.~\ref{fig:NondiagZetas}.
Consequently, for many phenomenological considerations the relations
\begin{eqnarray}\label{eq:c-to-a-approx}
   c_Z&=&c_W=a_V, \gap
   c_{\rho^0}=c_{\rho^\pm} = a_\rho = r a_V, 
   \nonumber\\ 
   c_{Z\rho^0}&=&c_{W\rho^\pm} = 0,
\end{eqnarray}
represent satisfactory approximations over quite a large region of $r$ values
and for all relevant values of $g''$ and $M_\rho$.

\begin{figure}
\centering
\includegraphics[scale=0.86]{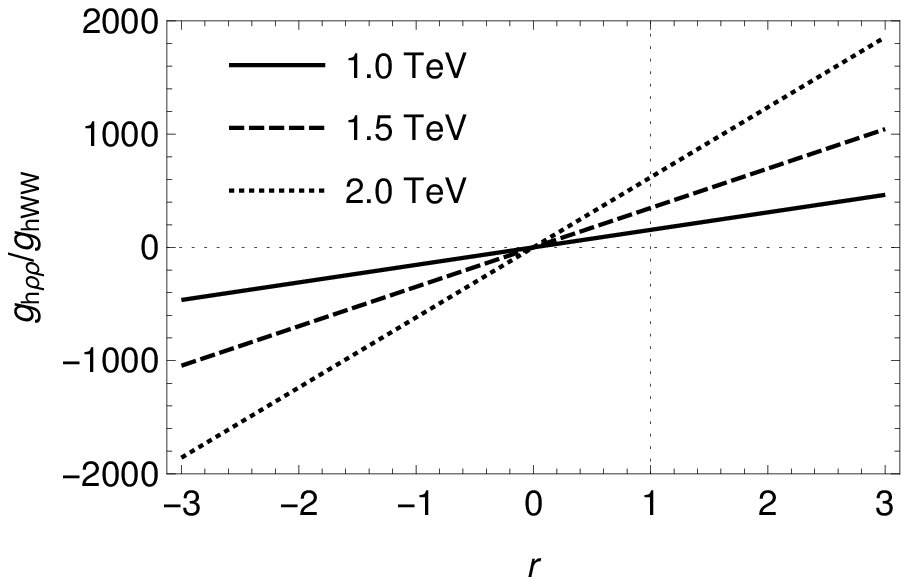}
\includegraphics[scale=0.86]{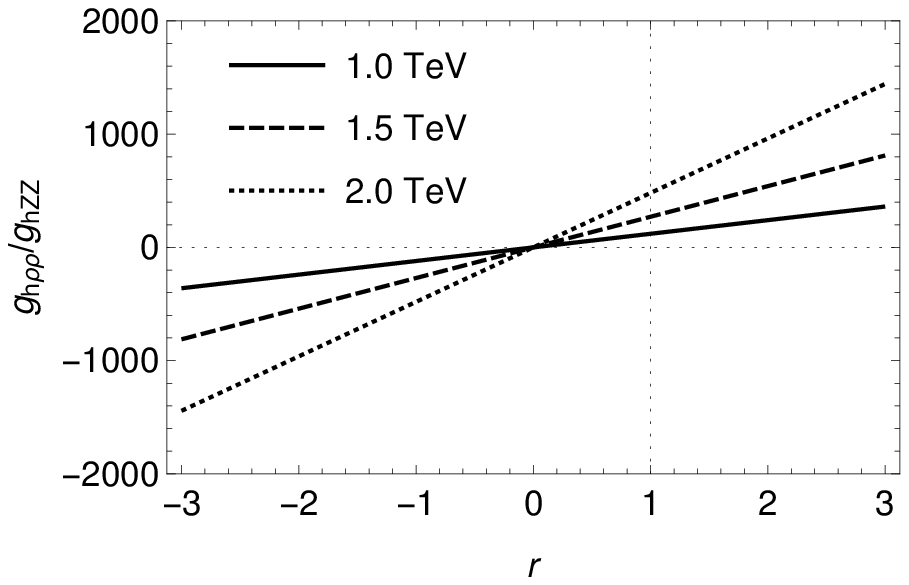}
\caption{\label{fig:RelativeStrengths1}
         The relative strengths
         of the $h\rho^+\rho^-$ and $h\rho^0\rho^0$ couplings 
         with respect to the $hW^+W^-$ and $hZZ$ couplings,
         respectively, as functions of the splitting $r=a_\rho/a_V$.
         The solid lines correspond to $M_\rho=1\unit{TeV}$,
         the dashed lines to $1.5\unit{TeV}$, and 
         the dotted lines to $2\unit{TeV}$.
         }
\end{figure}

Now, let us turn our attention to the Higgs-to-gauge-boson couplings.
Beside being proportional to the splitting factors the couplings
are amplified by the (squares of) gauge-boson masses. Namely,
\begin{equation}
   g_{hWW} = a_V\zeta_W M_W^2,\gap
   g_{hZZ} = a_V\zeta_Z M_Z^2/2,
\end{equation}
\begin{equation}
   g_{h\rho^+\rho^-} = a_V\zeta_{\rho^\pm} M_{\rho^\pm}^2,\gap
   g_{h\rho^0\rho^0} = a_V\zeta_{\rho^0} M_{\rho^0}^2/2,
\end{equation}
\begin{equation}
   g_{hW\rho^\pm} = a_V\zeta_{W\rho^\pm} M_W M_{\rho^\pm},\;
   g_{hZ\rho^0} = a_V\zeta_{Z\rho^0} M_Z M_{\rho^0}.
\end{equation}
Then
\begin{equation}\label{eq:ghWW-to-ghZZ}
   \frac{g_{hWW}}{g_{hZZ}} = 2\frac{\zeta_W}{\zeta_Z}\;\frac{M_W^2}{M_Z^2} =
   2[1+(r-1)\;\cO(x^4)]\;\frac{M_W^2}{M_Z^2}.
\end{equation}
Here, the whole dependence on new physics is contained in the $\cO(x^4)$ term. 
Thus, the splitting can modify the SM expectation for the ratio (\ref{eq:ghWW-to-ghZZ})
only very slightly.
On the other hand, owing to the new vector triplet's large mass new physics 
becomes manifest the most in the Higgs interaction with the new vector triplet
as can be seen in
\begin{eqnarray}
   \frac{g_{h\rho^+\rho^-}}{g_{hWW}} &=&
   \frac{\zeta_{\rho^\pm}}{\zeta_W}\;\frac{M_{\rho^\pm}^2}{M_W^2}
   \;\;\approx r\;\frac{M_{\rho^\pm}^2}{M_W^2},
   \\
   \frac{g_{h\rho^0\rho^0}}{g_{hZZ}} &=&
   \frac{\zeta_{\rho^0}}{\zeta_Z}\;\frac{M_{\rho^0}^2}{M_Z^2}
   \;\;\approx r\;\frac{M_{\rho^0}^2}{M_Z^2}.
\end{eqnarray}
Note that while the vector mass affects the ratios significantly
their dependences on $g''$ are completely ignorable. In Fig.~\ref{fig:RelativeStrengths1},
the ratios $g_{h\rho^+\rho^-}/g_{hWW}$ and $g_{h\rho^0\rho^0}/g_{hZZ}$
as functions of $r$ and for various vector resonance masses are depicted.

\begin{figure}
\centering
\includegraphics[scale=0.85]{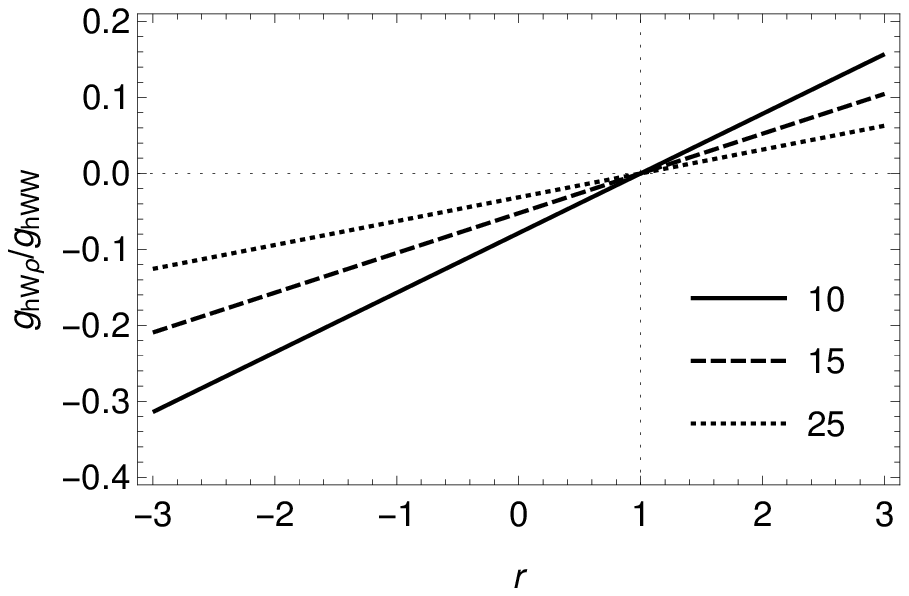}
\includegraphics[scale=0.85]{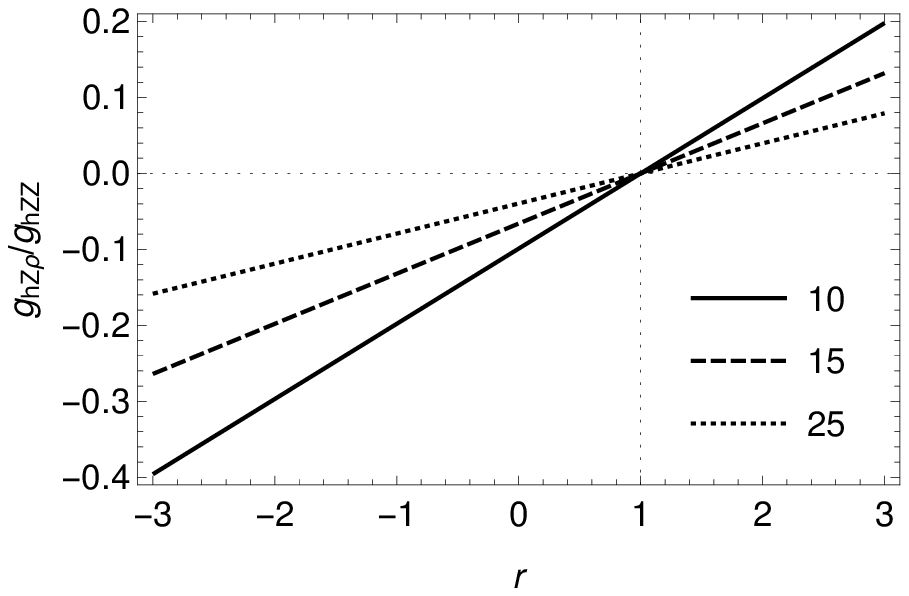}
\caption{\label{fig:RelativeStrengths2}
         The relative strengths of the $hW\rho^\pm$ and $hZ\rho^0$ couplings 
         with respect to the $hW^+W^-$ and $hZZ$ couplings,
         respectively, as functions of the splitting $r=a_\rho/a_V$.
         The solid lines correspond to $g''=10$,
         the dashed lines to $g''=15$, and 
         the dotted lines to $g''=25$.}
\end{figure}

If the splitting takes place new physics also manifests
via the emergence of two new vertices, $hW^\pm\rho^\pm$ and $hZ\rho^0$, 
not present either in the SM or in the Lagrangian $\Lagr_\mathrm{hV}$ when $a_\rho=a_V$.
Even though the new coupling strengths lag far behind the strengths of the $h\rho^+\rho^-$ and
$h\rho^0\rho^0$ couplings their presence would introduce new phenomena.
The relative strengths of the $hW^\pm\rho^\pm$ and $hZ\rho^0$ couplings with respect 
to the $hWW$ and $hZZ$ couplings, respectively, are given by 
\begin{eqnarray}
   \frac{g_{h W\rho}}{g_{hWW}} &=&
   \frac{\zeta_{W\rho}}{\zeta_W}\;\frac{M_{\rho^\pm}}{M_W}
   \;\; = \;\; (r-1)\;\cO(x),
   \\
   \frac{g_{h Z\rho}}{g_{hZZ}} &=&
   2\frac{\zeta_{Z\rho}}{\zeta_Z}\;
   \frac{M_{\rho^0}}{M_Z}\;\; = \;\; (r-1)\;\cO(x).
\end{eqnarray}
Note that while these ratios are affected significantly by $g''$
their dependences on the vector mass are completely ignorable. 
In Fig.~\ref{fig:RelativeStrengths2},
the ratios $g_{hW\rho^\pm}/g_{hWW}$ and $g_{hZ\rho^0}/g_{hZZ}$
as functions of $r$ and for various $g''$ values are depicted.

The splitting of $a_\rho$ from $a_V$ will also cause $c_W\neq c_Z$.
Thus, in principle, $r$ affects the model prediction for the custodial symmetry
protected rho parameter. Fortunately, the effect is negligible for a wide range of
$r$ values. To demonstrate it we inspect the prediction for the ratio
$\lambda_{WZ}\equiv c_W/c_Z$
\begin{equation}
   \lambda_{WZ} = \frac{\zeta_W}{\zeta_Z} = 1+(r-1)\;\cO(x^4).
\end{equation}
Numerically, our model predicts $|\lambda_{WZ}-1| < 6\times 10^{-5}$ at tree level
when $-3\leq r\leq 3$, $10\leq g''\leq 25$, and $M_\rho\in(1,2)\unit{TeV}$.
The current experimental limit is 
$\lambda_{WZ} = 0.88_{-0.09}^{+0.10}$~\cite{ATLAS-CMS-HiggsDataNew}.

%


\section{Limits on the Higgs-related parameters}
\label{sec:LimitsHiggsRelatedParams}

\subsection{Relevant measurements}
\label{subsec:relevant-measurements}
In this paper, we would like to identify the restrictions that the current
LHC measurements provide for the free parameters of the effective Lagrangian under
consideration. While the LHC cannot compete yet with the low-energy data from 
the LEP, SLC, and Tevatron in setting a limit on the value of $g''$, it certainly 
plays the key role in restricting the Higgs-related couplings of the EW gauge bosons,
fermions, and the vector resonance. 

In~\cite{tBESSepjc13}, we calculated the indirect limits
on $g''$ and other free parameters of the tBESS phenomenological Lagrangian,
a special case of the Lagrangian considered here,
fitting the observables $\epsilon_1$, $\epsilon_2$,
$\epsilon_3$, $\Gamma_b(Z\rightarrow b\bar{b}+X)$, and BR$(\bgs)$
under the assumption $a_V=a_\rho=1$ and $M_\rho=1\unit{TeV}$ and $2\unit{TeV}$.
In addition to the setup investigated here, the tBESS model contains three independent
direct interactions of the vector triplet with fermions: one with the left top-bottom quark 
doublet, another with the right top quark, and yet another with
the right bottom quark. The analysis of the limits lead to the conclusion
about the preference of higher $g''$ values, namely $g''>12$ at $95\%$~CL
when combined with preferred values of other free parameters. Further, 
the analysis seemed to support
the assumption of some models of partial fermion compositeness
that the new strong physics resonances couple
stronger to the right top quark than to the right bottom quark.

The requirement that our Lagrangian be treatable perturbatively bounds
the values of $g''$ from above by the naive perturbativity limit,
$g''/2\stackrel{<}{\sim} 4\pi$, implying $g''\stackrel{<}{\sim}25$.
If we took this value as the final say in this issue it would not be
reasonable to use $g''$ higher than about 20 in our calculations.
However, one can imagine that a more rigorous analysis of 
the perturbativity limit could somehow modify its value one way or the other. 

The limits on the Higgs-related couplings of the studied Lagrangian can be obtained
from the existing measurements of the LHC experiments. In this paper, we use constraints
from the ATLAS+CMS Collaborations analysis~\cite{ATLAS-CMS-HiggsDataNew}
based on the combination of six decay channels,
namely $H\rightarrow \gamma\gamma$, $ZZ^\ast$, $WW^\ast$, 
$b\bar{b}$, $\tau\bar{\tau}$ and $\mu\bar{\mu}$,
and of five production processes, namely gluon and vector boson fusions, 
and associated productions with $W$, $Z$, or a pair of top quarks.
In our analysis, we assume that there is only one SM-like Higgs boson state 
at about $125\unit{GeV}$ of a negligible decay width.

\subsubsection{The interim framework}

In~\cite{ATLAS-CMS-HiggsDataNew}, the combined ATLAS and CMS measurements were 
utilized to calculate fits and limits on the free kappa parameters of the interim framework
introduced in~\cite{kappa-framework}.
In this framework, kappas parameterize possible deviations of the Higgs interactions
from their SM expectations. The kappas are introduced so that
the squared free parameters $\kappa_i$ scale the SM Higgs production cross sections 
$\sigma_i^\mathrm{SM}$ and/or the partial Higgs decay widths $\Gamma_i^\mathrm{SM}$ 
associated with the $i$th SM particle.
When $\kappa_i=1$ then the best available SM predictions for $\sigma_i\cdot\mathrm{BR}_i$
are recovered so that for the true SM Higgs boson no artificial deviations caused by
ignored higher-order corrections are present.

The particular interim framework scenario that suits our effective Lagrangian
is that of Section~6.1 of~\cite{ATLAS-CMS-HiggsDataNew}. In this scenario, it is assumed 
that there are no non-SM decays of the Higgs
and that the branching ratio of invisible and/or undetected decay products is zero.
New particles in loops are allowed. Assuming that $\kappa_c = \kappa_t$, 
$\kappa_s=\kappa_b$, and $\kappa_\mu=\kappa_\tau$ there are generally seven free 
parameters in this scenario: five tree-level kappas
($\kappa_W,\kappa_Z,\kappa_t,\kappa_b,\kappa_\tau$) and
two loop-level kappas ($\kappa_g$ and $\kappa_\gamma$).
The loop-level kappas are associated with the higher-order effective vertices
$Hgg$ and $H\gamma\gamma$.

We will derive the limits on $a_V$, $a_\rho$, and $c_t$ fitting the experimental values
of kappas from Table 17 of \cite{ATLAS-CMS-HiggsDataNew}
(the $B_{\mathrm{BSM}}=0$ part).
The kappa values were obtained assuming $\kappa_t>0$ 
and using the following input parameters for 
the calculation of the SM quantities\footnote{Here we quote only the values of those input 
quantities that will also be needed as inputs in the calculations in this paper.}
~\cite{HandbookLHCHiggsXS1}: 
$M_W = 80.398\unit{GeV}$, $M_Z = 91.1876\unit{GeV}$,
$G_F=1.16637\times 10^{-5}\unit{GeV}^{-2}$, $M_t = 172.5\unit{GeV}$,
$M_h=125.5\unit{GeV}$. 
There are five kappas relevant for calculations of our limits: 
$\kappa_\gamma$, 
$\kappa_W$, $\kappa_Z$, $\kappa_g$, and $\kappa_t$. The remaining kappas,
$\kappa_b$ and $\kappa_\tau$ do not depend on
$a_V$, $a_\rho$, or $c_t$, at leading order.

The experimental uncertainties on kappas quoted in \cite{ATLAS-CMS-HiggsDataNew} are asymmetric.
In our analysis, we will simplify the situation by equalizing both sides of the uncertainty
to the one that is bigger. In addition, the 1-sigma interval of $\kappa_Z$ 
consists of two disconnected regions. In this case, we
take into account only the region where the best-fit value of $\kappa_Z$ is placed.
Thus modified experimental limits on $\kappa$'s that will be used in our calculations are shown 
in Table~\ref{tab:KappasFromATLAS}.

\begin{table}[h]
\centering
\caption{\label{tab:KappasFromATLAS}
         The experimental limits for $\kappa$'s that will be fitted to find the restrictions and 
         the best values of $a_V$, $a_\rho$, and $c_t$.
         Note that the kappa limits were obtained under the assumption that $\kappa_t> 0$.
        }
{\renewcommand{\arraystretch}{1.2}
\begin{tabular}{cc}
   \hline
   Parameter & ATLAS+CMS Limits \\
   \hline
   $\kappa_W$ & $\phantom{-}0.87{\pm 0.13}$ \\
   $\kappa_Z$ & $-0.98\pm 0.10$ \\
   $\kappa_t$ & $\phantom{-}1.40\pm 0.24$ \\
   $|\kappa_\gamma|$ & $\phantom{-}0.87\pm 0.14$ \\
   $|\kappa_g|$ & $\phantom{-}0.78\pm 0.13$ \\
   \hline
\end{tabular}}
\end{table}

\subsubsection{Fitting kappas with the effective Lagrangian parameters}

To restrict the free Higgs parameters via the fit to kappas the relations between 
the related kappas and $c_i$'s
have to be established. For that purpose we use some of the kappa defining 
quantities utilized in fitting the experiment and equate them to the predictions
of our effective theory. With the input parameters listed above we obtain
\begin{eqnarray}\label{eq:kappasWZt}
   \kappa_W^2&\equiv&\frac{\Gamma_{WW^*}}{\Gamma_{WW^*}^\mathrm{SM}}=c_W^2,\gap
   \kappa_Z^2\equiv\frac{\Gamma_{ZZ^*}}{\Gamma_{ZZ^*}^\mathrm{SM}}=c_Z^2,
   \nonumber\\
   \kappa_t^2&\equiv&\frac{\sigma_{t\bar{t}H}}{\sigma_{t\bar{t}H}^\mathrm{SM}}=c_t^2,\gap
   \kappa_g^2\equiv\frac{\sigma_{\mathrm{ggF}}}{\sigma_{\mathrm{ggF}}^\mathrm{SM}}=c_t^2,
\end{eqnarray}
where $\sigma_{t\bar{t}H}$ is the cross section of
associated production of the Higgs boson with a pair of top quarks,
$\sigma_{\mathrm{ggF}}$ is the gluon-fusion Higgs-production cross section,
and $\Gamma_{jj}$'s are the partial Higgs decay widths to the dibosons, $jj=WW^\ast, ZZ^\ast$. 
The index ``SM'' denotes the SM values.

As is well known, the $H\rightarrow\gamma\gamma$ decay in the SM occurs
at the loop level only.
In the SM, two dominant contributions originate from 
the Feynman diagrams with the $W$ boson and top quark loops.
Beyond the SM the anomalous couplings of the Higgs boson to $W$ and
top quark are parameterized by the factors $c_W$ and $c_t$, respectively.
In addition, the $H\rightarrow\gamma\gamma$ decay can be modified by the extra diagram
with the $\rho^\pm$ resonance in the loop. Thus, our effective theory predicts
\begin{equation}\label{eq:kappa_gamma}
   \kappa_\gamma^2\equiv\frac{\Gamma_{\gamma\gamma}}{\Gamma_{\gamma\gamma}^\mathrm{SM}}
   =\left[\frac{\alpha_\mathrm{EM}(g'', M_\rho)}{\alpha_\mathrm{EM}^\mathrm{SM}}\right]^2 
   \left[\frac{c_\gamma(c_W,c_t,c_{\rho^\pm})}{c_\gamma^\mathrm{SM}}\right]^2,
\end{equation}
where $c_\gamma^\mathrm{SM}$ is the SM coupling of the $H\gamma\gamma$ effective vertex
and $c_\gamma(c_W,c_t,c_{\rho^\pm})$ is its anomalous analog. Further,
$\alpha_\mathrm{EM}$ is the electromagnetic coupling constant which, in the case
of the effective Lagrangian and for the given set of the input 
parameters~\cite{HandbookLHCHiggsXS1}, depends on 
new physics. The leading order approximation that dominates the ratio 
of the squared $\alpha_\mathrm{EM}$'s when $M_\rho\gg M_Z$ reads
\begin{equation}\label{eq:RatioAlphasEM}
   \frac{\alpha^2_\mathrm{EM}(g'')}{{(\alpha_\mathrm{EM}^\mathrm{SM})}^2}
   = \frac{1}{ \left[1+16\sqrt{2}G_F M_W^2(1-M_W^2/M_Z^2)/g^{\prime\prime 2}\right]^2}.
\end{equation}
Numerically, when $g''$ varies between 10 and 25 then 
$(\alpha_\mathrm{EM}/\alpha_\mathrm{EM}^\mathrm{SM})^2$ 
changes from $0.992$ to $0.999$.
Thus, this effect will be ignored in our further analysis.
Also, note that in the leading order the ratio is not affected by the mass of the vector
resonance.

The contribution of the $\rho^\pm$ resonance to $h\rightarrow\gamma\gamma$ 
mimics the contribution of $W^\pm$; the only difference comes from different 
masses and couplings of the vector particles. 
Thus, considering the principal contributions only --- from the top, $W$, and $\rho$
loops --- $c_\gamma$ reads
\begin{eqnarray}\label{eq:cgamma}
   c_\gamma &=& \frac{1}{8} \left[c_t N_C q_t^2 F_\mathrm{ferm}(x_t)+
                               c_W F_\mathrm{vec}(x_W) \right.
   \nonumber\\ && \phantom{\frac{1}{8}[} 
		  \left.      +c_{\rho^\pm} F_\mathrm{vec}(x_\rho)\right],
\end{eqnarray}
where $N_C q_t^2=4/3$, $x_i=4M_i^2/M_h^2$, and
\begin{eqnarray}
   F_\mathrm{ferm}(x) &=& -2x[1+(1-x) f(x)], \\
   F_\mathrm{vec}(x) &=& 2+3x+3x(2-x)f(x),
\end{eqnarray}
where
\begin{equation}
    f(x) = \left\{
              \begin{array}{ll}
                 \arcsin^2(1/\sqrt{x}), & x\geq 1,\\
                 -\frac{1}{4}\left [\ln\left(\frac{1+\sqrt{1-x}}{1-\sqrt{1-x}}\right)-i\pi\right]^2, 
                 & x<1.
              \end{array}
           \right.
\end{equation}
For the given input values we get 
$F_\mathrm{ferm}(x_t) = -1.38$, $F_\mathrm{vec}(x_W)=8.34$, 
$F_\mathrm{vec}(x_{1\mathrm{TeV}})=7.01$, and $F_\mathrm{vec}(x_{2\mathrm{TeV}})=7.00$.
Then
\begin{eqnarray}\label{eq:cgammaMrho1-2TeV}
   c_\gamma(M_\rho = \mbox{1--2}\unit{TeV}) &=&
   -0.23\;c_t +1.04 \;c_W+0.88\; c_{\rho^\pm}
   \nonumber\\
   &=& -0.23\;c_t +a_V\;[1.04\;\zeta_W(r)
   \nonumber\\ && \phantom{-0.23\;c_t +a_V}
   +0.88\;\zeta_{\rho^\pm}(r)].
\end{eqnarray}
Note that, for the displayed decimal places, the numerical coefficients in this formula are not 
sensitive to varying the $\rho$ mass between 1 and 2~TeV.

The SM value of $c_\gamma$ is obtained when $c_t=c_W=1$ and $c_{\rho^\pm}=0$: 
$c_{\gamma}^\mathrm{SM}=0.81$.

The effective Lagrangian predictions for observables do not change under the 
simultaneous sign change of all $c_i$ parameters. Therefore, a sign of
one of $c_i$'s can be fixed without losing physically distinguishable configurations of theory.
Owing to that we choose $c_t>0$ throughout the paper.


\subsection{Failure of the \bm{$a_V=a_\rho$} scenario}
\label{subsec:aV-equals-arho}

\begin{figure}
\centering
\includegraphics[scale=0.90]{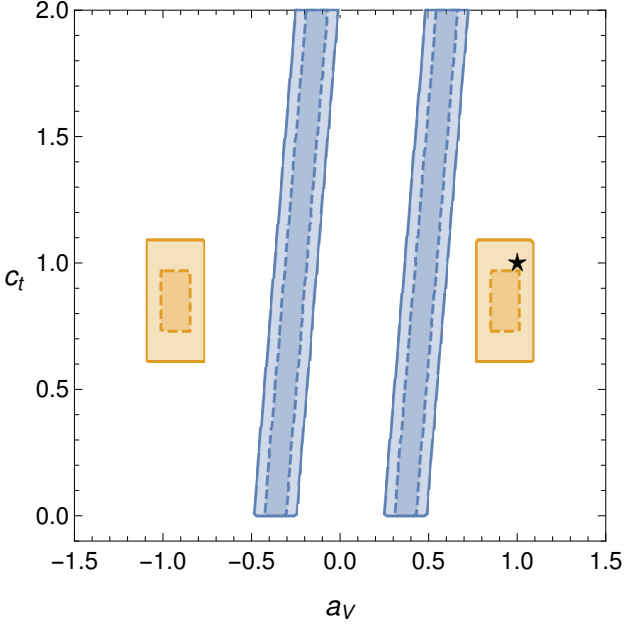}
\caption{\label{fig:Hgammagamma_aV_ct}
         The 1-sigma (dashed) and 2-sigma (solid) allowed regions (blue stripes) 
         in the $a_V-c_t$ plane
         derived from the experimental restriction on $\kappa_\gamma$ under
         the no-splitting assumption. The orange rectangular areas correspond to the 1-sigma
         and 2-sigma allowed regions for $a_V$ and $c_t$ 
         obtained by separate fitting of
         $\kappa_W^2, \kappa_Z^2$ and $\kappa_t^2, \kappa_g^2$, respectively.
         The star indicates the SM values.
         }
\end{figure}

Before performing the full fit on $\kappa_\gamma, \kappa_W, \kappa_Z, \kappa_g$, and
$\kappa_t$ with three free parameters $a_V, a_\rho, c_t$
let us provide a simple demonstration that the no-splitting scenario,
$a_V=a_\rho$, has a hard time to satisfy the experimental restrictions on the 
kappas under consideration.
Applying (\ref{eq:kappa_gamma}) to this situation we obtain
\begin{equation}\label{eq:kappa_gamma_nosplitting}
   |\kappa_\gamma| = |2.37\;a_V-0.28\;c_t|.
\end{equation}
Using the experimental restriction on $\kappa_\gamma$ (see Table~\ref{tab:KappasFromATLAS})
the Eq.~(\ref{eq:kappa_gamma_nosplitting}) results in the allowed region comprised of two
parallel stripes crossing the $a_V-c_t$ plane as shown in Fig.~\ref{fig:Hgammagamma_aV_ct}.
Let us recall that we work under the assumption $c_t>0$ which reduces the $a_V-c_t$ plane to
a half-plane.

In addition, the graphs in Fig.~\ref{fig:Hgammagamma_aV_ct} 
display rectangular intersections of the experimentally allowed regions for $a_V$ and $c_t$. 
The limit on $a_V$ is derived by fitting $\kappa_W^2$ and $\kappa_Z^2$
while taking into account the no-splitting relations $c_W=c_Z=a_V$.
The relevant $\chi^2$-function consists of the last two terms 
of~(\ref{eq:chi2squaredkappas-approx}). The obtained fit reads
$|a_V|=0.93\pm0.08$ where the absolute value originate in the fitting of squares
of variables. Analogically, the $c_t$ constraint is obtained by fitting
$\kappa_t^2$ and $\kappa_g^2$. The corresponding $\chi^2$-function
contains the second and third terms of~(\ref{eq:chi2squaredkappas-approx}).
The obtained best value of $|c_t|$ is $c_t=0.85^{+0.11}_{-0.12}$.

The graphs in Fig.~\ref{fig:Hgammagamma_aV_ct} indicate a tension 
between the $h\rightarrow\gamma\gamma$ limits and the combined 
$\kappa_W-\kappa_Z$ and $\kappa_t-\kappa_g$ restrictions
in the no-splitting version of our effective Lagrangian.
In Section~\ref{subsec:CombinedAnalysis}, we will assess if and to what extent  
this tension can be 
removed by allowing $a_V$ and $a_\rho$ to become independent.


\subsection{Full fit analysis}
\label{subsec:CombinedAnalysis}

Assuming that $a_V$ and $a_\rho$ are free independent parameters of 
the effective Lagrangian under consideration, we calculate limits on 
$a_V$, $a_\rho$, and $c_t$ and search for their best values 
minimizing the $\chi^2$-function built out of the relevant measured observables 
and their predictions.
We use the experimental values of $\kappa_W$, $\kappa_Z$, $\kappa_t$, $\kappa_\gamma$, 
and $\kappa_g$
shown in Table~\ref{tab:KappasFromATLAS}. Theoretical predictions of the kappas 
in terms of $c_Z$, $c_W$, $c_{\rho^\pm}$, and $c_t$ of the Lagrangian interaction
term (\ref{eq:HcplngLinear}) and (\ref{eq:LagFermScalar}), respectively,
are given in (\ref{eq:kappasWZt}) and (\ref{eq:kappa_gamma}). 
The dependence of $c_\gamma$ on $c_W$, $c_{\rho^\pm}$, and $c_t$
in (\ref{eq:kappa_gamma}) is given by (\ref{eq:cgamma}).
The $c_Z$, $c_W$, and $c_{\rho^\pm}$ are, in turn, 
related to $a_V$ and $a_\rho$ via (\ref{eq:cecka}).

In the interim framework the squares of kappas rather than the kappas themselves are
defined and related to measured observables. Therefore, our theoretical predictions
are also related to squared kappas and the best fit should be sought for minimizing 
the following $\chi^2$-function:
\begin{eqnarray}\label{eq:chi2squaredkappas}
   \chi^2(a_V,a_\rho,c_t) &=&
   \left\{\frac{\kappa_\gamma^2-[c_\gamma(a_V,a_\rho,c_t)/c_\gamma^\mathrm{SM}]^2}
   {\tilde{\sigma}_{\gamma}}\right\}^2
   \nonumber\\ &&
   +\sum_{i=t,g}
   \left(\frac{\kappa_i^2-c_t^2}{\tilde{\sigma}_i}\right)^2
   \nonumber\\ &&
   +\sum_{i=W,Z}
   \left[\frac{\kappa_i^2-c_i^2(a_V,a_\rho)}{\tilde{\sigma}_{i}}\right]^2,
\end{eqnarray}
where $\tilde{\sigma}_i$'s are the experimental errors for $\kappa_i^2$'s.
The fitting of squares introduces degeneracy of solutions caused by
the insensitivity of $\chi^2$ to
the relative signs between the kappas and theory parameters. Recall that $c_t>0$ by assumption.

If we approximate $c_W=c_Z=a_V$ and $c_\rho^\pm=a_\rho$
the Eq.~(\ref{eq:chi2squaredkappas}) can be simplified 
without any significant impact on the best-fit values
as can be inferred from our conclusions obtained in 
Section~\ref{sec:efflagr} (viz., the Eq.~(\ref{eq:c-to-a-approx})). 
Then, substituting 
(\ref{eq:cgammaMrho1-2TeV}) into (\ref{eq:chi2squaredkappas}) we obtain
\begin{eqnarray}\label{eq:chi2squaredkappas-approx}
   \tilde{\chi}^2(a_V,a_\rho,c_t)&=&
   \left[\frac{\kappa_\gamma^2-(1.28\; a_V+1.09\; a_\rho -0.28\; c_t)^2}
              {2\kappa_\gamma\sigma_\gamma}\right]^2
   \nonumber\\
   && +\left(\frac{\kappa_t^2-c_t^2}{2\kappa_t\sigma_t}\right)^2
   +\left(\frac{\kappa_g^2-c_t^2}{2\kappa_g\sigma_g}\right)^2
   \nonumber\\
   && +\left(\frac{\kappa_W^2-a_V^2}{2\kappa_W\sigma_W}\right)^2
      +\left(\frac{\kappa_Z^2-a_V^2}{2\kappa_Z\sigma_Z}\right)^2.
\end{eqnarray}
It is obvious that $\tilde{\chi}^2$ has degenerate minima. If
the $\tilde{\chi}^2$-function assumes its minimum value $\tilde{\chi}^2_\mathrm{min}$
for some triplet $a_V,a_\rho,c_t$ 
then also $\tilde{\chi}^2(-a_V,-a_\rho,-c_t)=\tilde{\chi}^2_\mathrm{min}$.
Nevertheless, by fixing $c_t>0$ we eliminate a half of the degenerate minima.

A simple inspection of the Eq.~(\ref{eq:chi2squaredkappas-approx}) implies that 
$\tilde{\chi}^2$ can be minimized when, at the same time, 
$c_t$ assumes the value of ``weighted average'' of $\kappa_t$ and $\kappa_g$,
$|a_V|$ assumes a ``weighted average''
of $\kappa_W$ and $\kappa_Z$, and $a_\rho$ sets to zero
the first term with the before-obtained values of $c_t$ and $a_V$ substituted in.
Zeroing the first term of ~(\ref{eq:chi2squaredkappas-approx}) amounts to
solving a quadratic equation in $a_\rho$. Thus, there are generally two solutions
for $(a_V, c_t)$ and two solutions for $(-a_V, c_t)$, i.e. four solutions in total.
Since the first term of Eq.~(\ref{eq:chi2squaredkappas-approx}) has zero contribution
to $\tilde{\chi}^2_\mathrm{min}$ and because the following two terms
depend only on $c_t$ and two last terms only
on $|a_V|$, all four solutions result in the same value of $\tilde{\chi}^2_\mathrm{min}$.

In particular, by fitting the kappa values of Table~\ref{tab:KappasFromATLAS}
we get four minimizing triplets of $\{a_V,a_\rho,c_t\}$ with the same minimal
values of $\tilde{\chi}^2$, $\tilde{\chi}^2_\mathrm{min}=4.17$. Having $\mbox{d.o.f.}=5-3=2$,
the value corresponds to the hypothesis backing of $12\%$. The values of the degenerate 
best-fit triplets along with the corresponding parameter constraints 
at 20, 68, and $95\%$~CL are shown 
in Table~\ref{tab:numericalresults}.

\begin{table*}[t]
\centering
\caption{\label{tab:numericalresults}
         The best-fit values of $c_t$, $a_V$, and $a_\rho$ corresponding to the four minima
         of the $\tilde{\chi}^2$-function (\ref{eq:chi2squaredkappas-approx})
         with the constraints at $20\%$~CL (1-sigma), $68\%$~CL, and $95\%$~CL. 
         All minima (labeled as A, B, C, and D) have the same backing of $12\%$.
        }
{\renewcommand{\arraystretch}{1.5}
\begin{tabular}{|cc|cc|cc|cc|cc|}
   \hline
   \multicolumn{2}{|c|}{Parameter} & \multicolumn{2}{c|}{A} & \multicolumn{2}{c|}{B} & 
                                          \multicolumn{2}{c|}{C} & \multicolumn{2}{c|}{D} \\
   \hline
    & \mbox{$20\%$~CL} & & $^{+0.11}_{-0.12}$ & & $^{+0.11}_{-0.12}$ & & $^{+0.11}_{-0.12}$ & & $^{+0.11}_{-0.12}$ \\
   $c_t$ & \mbox{$68\%$~CL} & 0.85 & $^{+0.19}_{-0.25}$ & 0.85 & $^{+0.19}_{-0.25}$ & 0.85 & $^{+0.19}_{-0.25}$ & 0.85 & $^{+0.19}_{-0.25}$ \\
    & \mbox{$95\%$~CL} & & $^{+0.27}_{-0.43}$ & & $^{+0.27}_{-0.43}$ & & $^{+0.27}_{-0.43}$ & & $^{+0.27}_{-0.43}$ \\
   \hline
    & \mbox{$20\%$~CL} & & $^{+0.08}_{-0.08}$ & & $^{+0.08}_{-0.08}$ & & $^{+0.08}_{-0.08}$ & & $^{+0.08}_{-0.08}$ \\
   $a_V$ & \mbox{$68\%$~CL} & 0.93 & $^{+0.14}_{-0.16}$ & 0.93 & $^{+0.14}_{-0.16}$ & -0.93 & $^{+0.16}_{-0.14}$ & -0.93 & $^{+0.16}_{-0.14}$ \\
    & \mbox{$95\%$~CL} & & $^{+0.20}_{-0.25}$ & & $^{+0.20}_{-0.25}$ & & $^{+0.25}_{-0.20}$ & & $^{+0.25}_{-0.20}$ \\
   \hline
    & \mbox{$20\%$~CL} & & $^{+0.16}_{-0.17}$ & & $^{+0.17}_{-0.15}$ & & $^{+0.17}_{-0.16}$ & & $^{+0.16}_{-0.17}$ \\
   $a_\rho$ & \mbox{$68\%$~CL} & -0.08 & $^{+0.29}_{-0.34}$ & -1.68 & $^{+0.34}_{-0.28}$ & 0.52 & $^{+0.33}_{-0.29}$ & 2.11 & $^{+0.28}_{-0.34}$ \\
    & \mbox{$95\%$~CL} & & $^{+0.43}_{-0.58}$ & & $^{+0.58}_{-0.40}$ & & $^{+0.58}_{-0.44}$ & & $^{+0.41}_{-0.58}$ \\
   \hline
\end{tabular}}
\end{table*}

\begin{figure}
\centering
\includegraphics[scale=0.85]{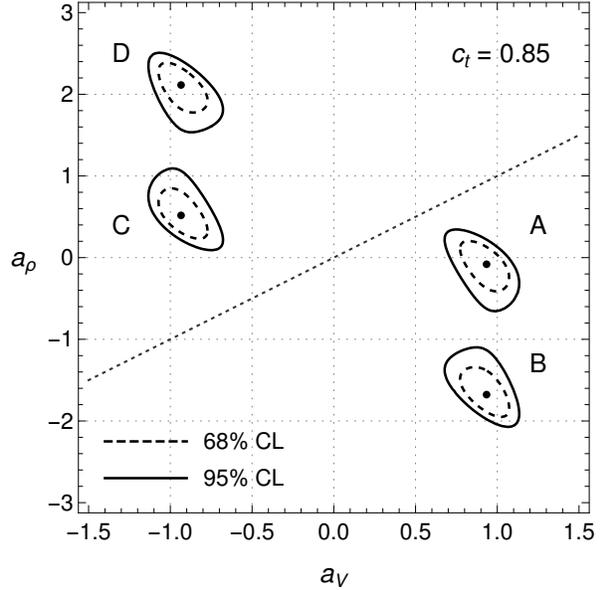}
\caption{\label{fig:chi2tilda_contours}
	   The two-dimensional cut 
	   of the three-dimensional region of the allowed values of parameters 
	   $a_V$, $a_\rho$, and $c_t$ 
	   when $c_t$ is fixed at its best value of $0.85$.
	   The dots indicate the best-fit values of the fit.
	   They are labeled as A, B, C, and D in correspondence with
	   Table~\ref{tab:numericalresults}. The dashed and solid contours
	   show $68\%$~CL and $95\%$~CL limits.
	   The straight line indicates the points of the no-splitting scenario, $a_\rho=a_V$.
	   }
\end{figure}

\begin{figure}
\centering
\includegraphics[scale=0.85]{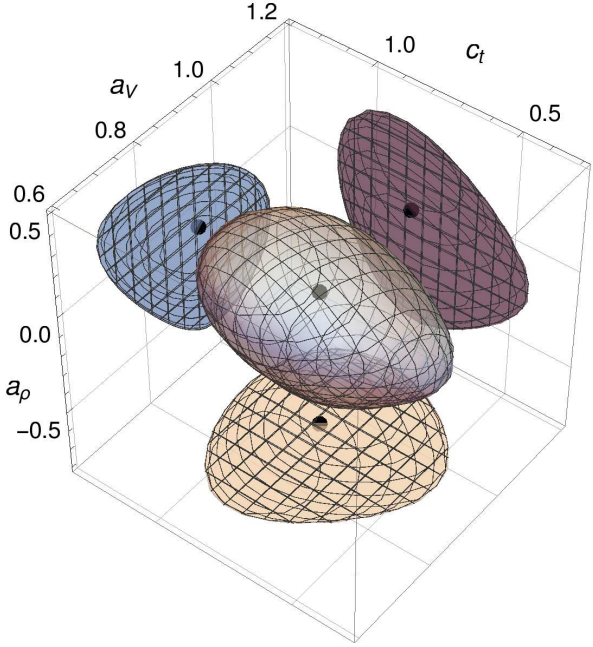}
\caption{\label{fig:4x3D}
         The $95\%$~CL allowed region of the parameters $a_V$, $a_\rho$, and $c_t$ 
         of the $\tilde{\chi}^2$ minima A 
         and the region projections to two-parameter planes.
         Dots indicate the best-fit value and its projections to the planes.         
	 }
\end{figure}

The graphic representation of the best-fit values is depicted in Fig.~\ref{fig:chi2tilda_contours}.
There, the two-dimensional cut of the $c_t-a_\rho-a_V$ allowed regions by the $c_t=0.85$ plane
is shown. Note that $0.85$ is the best-fit value of $c_t$. The contours depicted in the graph
correspond to the 68 and $95\%$~CL regions in the $c_t - a_\rho - a_V$ space.
The splitting factors $r$ for the best-fit points A, B, C, and D have the values $-0.09$,
$-1.81$, $-0.56$, and $-2.27$, respectively.
The straight line indicates the points of the no-splitting scenario, $a_\rho=a_V$.
The full $95\%$~CL region in the $c_t - a_\rho - a_V$ space around the best-fit point A
is shown in Fig.~\ref{fig:4x3D}. The allowed regions around the best-fit points B, C, and
D are very similar in shape and size to the region A.

There is a good reason why $a_V a_\rho<0$ for all four best-fit points. It is because the 
combined contributions of $a_V$ and $c_t$ to $|c_\gamma|$ --- the values of $a_V$ and $c_t$
being determined by the other terms of the $\chi^2$-function~(\ref{eq:chi2squaredkappas-approx})
--- overshoot the optimal value of $|c_\gamma|$. Since the $a_V$ contribution dominates
the $c_t$ contribution, the $a_\rho$ has to have a sign opposite to the sign of $a_V$
in order to counterbalance the excess. In addition, since we optimize 
$|c_\gamma|$, rather than $c_\gamma$ itself, $a_\rho$ resulting in the optimal $c_\gamma$
plays as well as $a_\rho$ resulting in $-c_\gamma$. Thus, we end up with two $a_\rho$'s for
each of the two best-fit values of $a_V$. 

It is not a difficult exercise to impose the no-splitting condition, $a_\rho=a_V$,
on $\tilde{\chi}^2$ of (\ref{eq:chi2squaredkappas-approx}) in order to obtain a more
rigorous quantification and justification of the conclusion we have reached 
in Section~\ref{subsec:aV-equals-arho}.
The fitting of $\tilde{\chi}^2$ results in two minima of unequal depths.
This asymmetry results from the sensitivity of $\tilde{\chi}^2$ on 
the relative sign of $a_V$ and $c_t$ in the $\kappa_\gamma$ term.
The global minimum assumes the value $\tilde{\chi}_\mathrm{min}^2 = 21.3$ which for 
$\mbox{d.o.f.}=5-2=3$ corresponds to the hypothesis backing of $0.9\%$.
The minimum resides at $a_V=0.55\pm0.05$ and $c_t=0.89^{+0.10}_{-0.11}$.
The local minimum at $(a_V, c_t)=(-0.33,0.81)$ amounts to the value higher 
by $\Delta\tilde{\chi}^2=10.7$ above the global minimum. 

\begin{figure}
\centering
\includegraphics[scale=0.85]{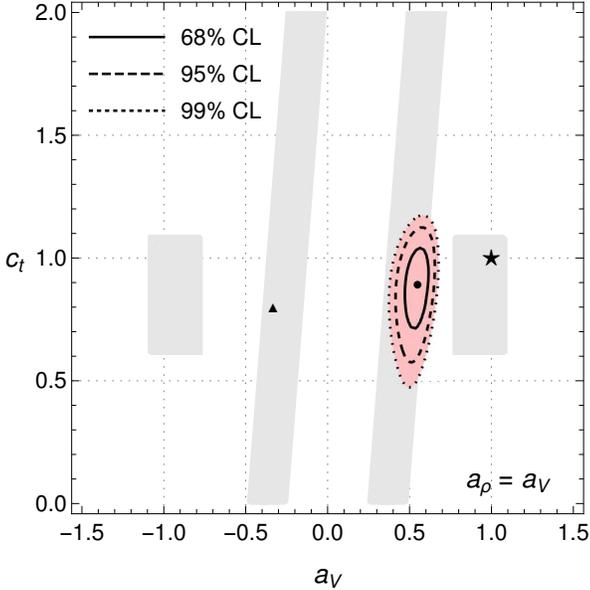}
\caption{\label{fig:no-splitting_fits_together}
         The fit under the no-splitting condition and the related CL contours.
         The multiply encircled red area corresponds to 
         the allowed regions of 68, 95, and
         $99\%$~CL. The dot inside the area indicates the best-fit value.
         The triangle indicates the position of the second (local) minimum
         of the $\chi^2$-function. The star indicates the SM values of
         the fitting parameters. For comparison, 
         the graph is interlaced with the reminiscence
         of Fig.~\ref{fig:Hgammagamma_aV_ct} (gray regions) where the partial limits
         obtained by a cruder approach were shown.
         }
\end{figure}

The fit under the no-splitting condition and the related CL
contours are depicted in Fig.~\ref{fig:no-splitting_fits_together}.
Besides, the graph contains reminiscence of Fig.~\ref{fig:Hgammagamma_aV_ct}
where the limits obtained by the cruder approach of Section~\ref{subsec:aV-equals-arho}
were shown. From a naive visual inspection we would say that the results of the cruder
analysis do not contradict the more sophisticated fit performed here. Not only the position
of the minimum coincides with the guessed expectation based on Fig.~\ref{fig:Hgammagamma_aV_ct},
but the obtained hypothesis backing
also confirms our qualitative conclusion of Section~\ref{subsec:aV-equals-arho} 
about a low support of the data for the no-splitting scenario.

Having the rho-to-Higgs coupling disconnected from the Higgs interactions with the EW 
gauge bosons and acknowledging that $a_V=c_t=1$ is compatible with the experiment,
one might wonder how successful the parameter $a_\rho$ alone would be in accommodating 
the existing data if $a_V$ and $c_t$ were kept on their SM values. 
This simple exercise amounts to minimizing 
the $\tilde{\chi}^2$-function~(\ref{eq:chi2squaredkappas-approx}) when $a_V=c_t=1$.
There are two minima, at $a_\rho=-0.12$ and $a_\rho=-1.72$, both with 
$\chi^2_\mathrm{min}=7$. Since $\mbox{d.o.f.}=5-1=4$ our effective Lagrangian with
the SM couplings of the Higgs boson to the EW gauge bosons and top quark has
$13.8\%$ backing by the data. Raising $\chi^2$ above its minimum by 1 
the 1-sigma ($68\%$~CL) constraints read $^{+0.12}_{-0.14}$ and $^{+0.14}_{-0.12}$
for the former and latter $a_\rho$ best values, respectively. Hence, if we had
a theoretical reason to demand $a_V=c_t=1$ the experimental support for the vector triplet
of particular $a_\rho$ values would be as good as in the model with loose $a_V$ and $c_t$.

There are two kappas in the combined ATLAS+CMS measurements~\cite{ATLAS-CMS-HiggsDataNew}
not utilized in our analysis: 
$\kappa_\tau\equiv\Gamma_{\tau^+\tau^-}/\Gamma_{\tau^+\tau^-}^\mathrm{SM}$ and 
$\kappa_b\equiv\Gamma_{b\bar{b}}/\Gamma_{b\bar{b}}^\mathrm{SM}$.
In the approximation used in this paper, $\kappa_\tau$ is solely related to $c_\tau$,
$\kappa_\tau^2=c_\tau^2$. Thus, it has no impact on the fits of other parameters.
Regarding $\kappa_b$, while it shares the dependence on $c_b$ with $\kappa_g$, its effect 
on the Higgs production via gluon-gluon fusion is negligible. Consequently, we had dropped
the $\kappa_b$ term from the $\chi^2$-function.

On the other hand, the effective coupling 
$\kappa_{Z\gamma}\equiv\Gamma_{Z\gamma}/\Gamma_{Z\gamma}^\mathrm{SM}$
has a similar structure as $\kappa_\gamma$: it depends on $a_V$, $a_\rho$, and $c_t$ 
parameters at leading order. Thus, it has a potential to affect our fits significantly.
Unfortunately, the existing measurements restrict $\kappa_{Z\gamma}$ very 
poorly~\cite{ATLAS-HiggsData}. Because of that, neither $\kappa_{Z\gamma}$ was included
in the $\chi^2$-function~(\ref{eq:chi2squaredkappas}).

\subsection{Unitarity limits for the preferred values of the Higgs couplings}
\label{subsec:UnitarityLimits}

In this Section, we would like to determine how the usability of
our phenomenological Lagrangian is restricted by the unitarity limits
when the data preferred Higgs couplings obtained 
in Subsection~\ref{subsec:CombinedAnalysis}
are considered.

Opposite to the SM, our effective Lagrangian is not renormalizable
and its applicability is limited to a finite range of energies,
not exceeding the point where scattering amplitudes violate
unitarity. Considering the low-energy phenomenology of underlying 
new strong physics it is natural to demand that its successful
effective description does not break down below the scale significant
to these new interactions. Thus, the results of the investigation 
of the unitarity limits could be suggestive either of the new physics
scale or, in a less fortunate case, 
of the defects in our Lagrangian. The least we can, and need, to
deduce from such an analysis is the range of usability of the effective
Lagrangian we play with.

The unitarity of our Lagrangian was investigated in detail in our previous 
publications~\cite{tBESSprd11,tBESSepjc13}. Our analysis was based on
the scattering amplitudes of the longitudinal EW gauge bosons to the EW
gauge bosons.
We studied the unitarity of the amplitudes using the Equivalence theorem
approximation where the concerned amplitudes were replaced by the pionic
scattering amplitudes of the non-linear sigma model. While the approximation
corresponds to the limit $g,g'\rightarrow 0$ (no pion to EW gauge-boson vertices), 
the exchange of the Higgs and vector resonances was included.

Even more thorough investigation of the unitarity limits of the same (in relevant
sectors) effective 
Lagrangian was performed in~\cite{CompositeHiggsSketch}. Beside the elastic
$\pi\pi\rightarrow\pi\pi$ amplitudes, authors of~\cite{CompositeHiggsSketch}
analyzed the unitarity limits implied by the non-elastic 
$\pi\pi\rightarrow hh, \rho_L\rho_L, h\rho_L$ processes.
Note that wherever applicable the conclusions of~\cite{CompositeHiggsSketch}
coincide with our conclusions~\cite{tBESSprd11,tBESSepjc13}.
In the following, we utilize the findings of these papers for calculation
of the tree-level unitarity limits for our Lagrangian. Skipping details of 
the very standard calculations, below we summarize and discuss the obtained results.

\begin{figure}
\centering
\includegraphics[scale=0.85]{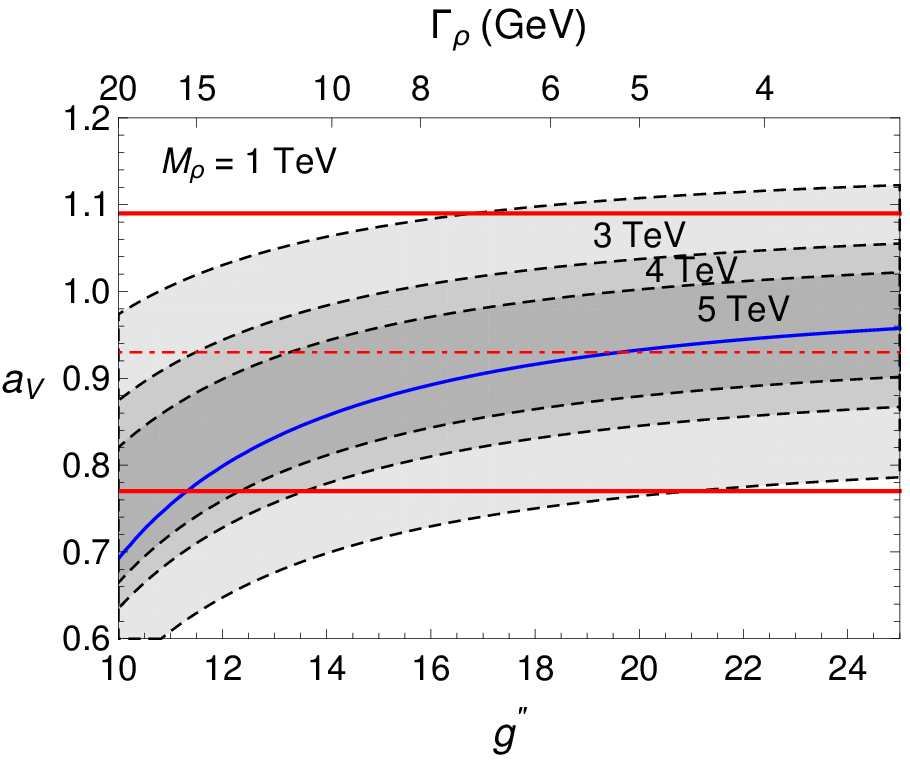}\vspace{2mm}
\includegraphics[scale=0.85]{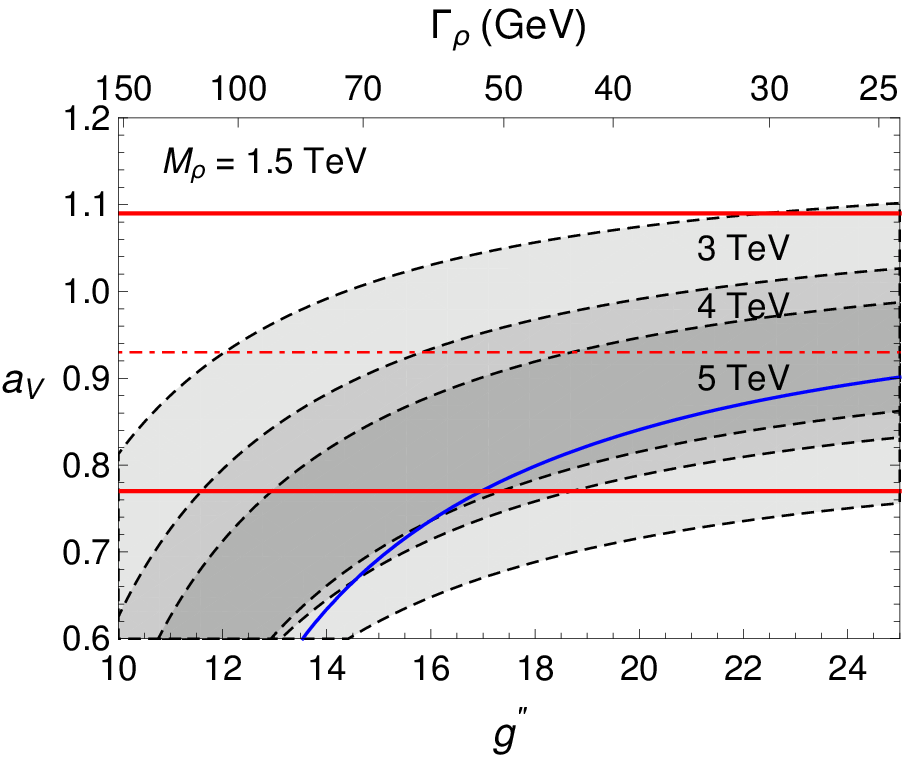}\vspace{2mm}
\includegraphics[scale=0.85]{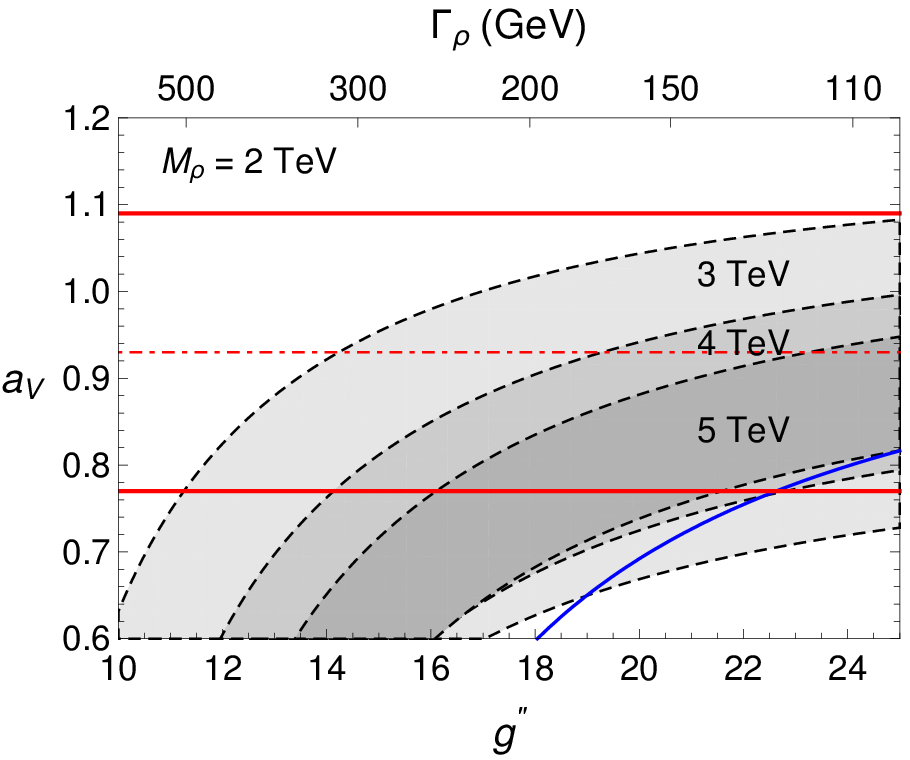}
\caption{\label{fig:U-limits}
         The tree-level unitarity constraints from $\pi\pi\rightarrow\pi\pi$
         in the $a_V-g''$ plane for different masses of the vector triplet,
         $M_\rho=1$, $1.5$, and $2\unit{TeV}$ in the clockwise direction. 
         The shaded areas indicate regions where the unitarity holds, up to 3, 4, and 5~TeV.
         Two horizontal (solid red) lines show allowed interval of values of $a_V$
         at 2-sigma, the dot-dashed line is the best-fit value $a_V=0.93$.
         The blue solid curve plots the sum rule~(\ref{eq:SumRule-pipipipi}).
         The upper axis labels show the vector resonance decay widths corresponding
         to $g''$'s of the bottom axis.
         }
\end{figure}

In Fig.~\ref{fig:U-limits}, we show the $\pi\pi\rightarrow\pi\pi$ tree-level
unitarity limits\footnote{
All our calculations use the $\Gamma_\rho=0$ approximation. This is justifiable 
as long as $M_\rho$ is far below the unitarity limit in terms of the $\rho$ width.}. 
In our case, the $\pi\pi\rightarrow\pi\pi$ scattering amplitudes depend only on
one of the three Higgs couplings investigated in this paper, namely $|a_V|$.
Besides, the amplitudes also depend on $M_\rho$ and $g''$. 
There are three graphs in Fig.~\ref{fig:U-limits} corresponding to three
different masses of the vector resonance:
$M_\rho = 1$, 1.5, and $2\unit{TeV}$. In each graph,
there are regions shown where unitarity holds up to $\Lambda=3\unit{TeV}$, 
$4\unit{TeV}$, and $5\unit{TeV}$. The regions are superimposed by the experimentally allowed
2-sigma interval for $a_V$ obtained in Subsection~\ref{subsec:CombinedAnalysis}: 
$0.77\leq a_V\leq 1.09$, the best fit being $a_V=0.93$.

We can see that for the $1\unit{TeV}$ vector resonance unitarity holds
up to at least $\Lambda=3\unit{TeV}$ when $a_V$ and $g''$ assumes allowed 
values\footnote{Recall that $12\leq g''\leq 8\pi$.}, excluding a small region
where $g''\rightarrow 12$ and $a_V\rightarrow 1.1$. When raising $M_\rho$,
the region where unitarity holds shrinks toward higher $g''$. At the same time,
it slightly shifts toward smaller $a_V$.

\begin{figure}
\centering
\includegraphics[scale=0.87]{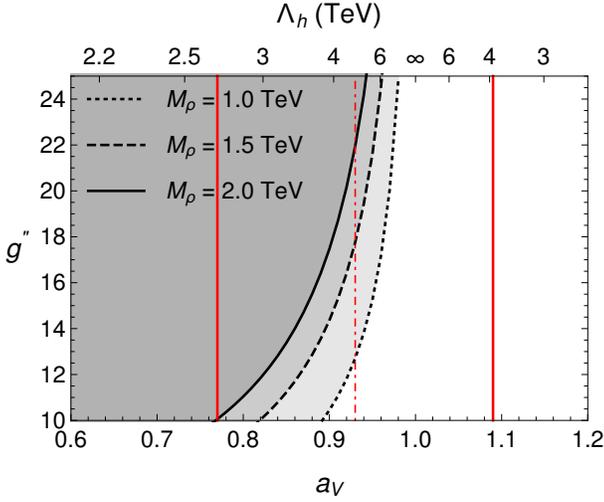}
\caption{\label{fig:HvsRho}
         The regions where the unitarity limit of our model ($a_V,g'',M_\rho$) gets bigger
         (darker area) and smaller (lighter area) than the unitarity limit $\Lambda_h(a_V)$ 
         of the Higgs-only model of a given value $a_V$. The regions are depicted for three
         different masses $M_\rho=1$, $1.5$, and $2\unit{TeV}$.
         Two vertical (solid red) lines show allowed interval of values of $a_V$
         at 2-sigma, the dot-dashed line is the best-fit value $a_V=0.93$.
         }
\end{figure}

If we dropped the vector resonance from our effective Lagrangian 
the Higgs resonance alone could unitarize the elastic $\pi\pi$ amplitudes up to 
some $\Lambda_h$ that depends on $a_V$; 
for example, if $a_V$ assumes its best value of $0.93$ the tree-level unitarity holds up to 
$\Lambda_h=4.6\unit{TeV}$. Perhaps, it might also be worth mentioning that 
$\Lambda_h\rightarrow\infty$ when $a_V=1$.
One wonders how adding the vector triplet to the Higgs-only
setup changes the $\Lambda_h$ limit. To assess it we plot 
the graph in Fig.~\ref{fig:HvsRho}. There, for given $a_V$ and $M_\rho$, 
the values of $g''$ are divided into two intervals: the one where $\Lambda<\Lambda_h(a_V)$, and
the other, where $\Lambda>\Lambda_h(a_V)$. 
We can see in the graph that if $a_V\geq 1$
adding the vector resonance will always lower the unitarity limit.
On the other hand, if $a_V< 1$ there is always $g_0''$ such that for all
$g''\geq g_0''$ the unitarity limit gets bigger.

In general, the $\pi\pi$ scattering amplitudes are plagued with the linear 
growths in $s$. Any added ingredient or assumption that removes the linear
growths has a good chance to improve unitarity limits. 
It can be shown~\cite{CompositeHiggsSketch} that
assuming special relations among the parameters of our Lagrangian (the sum rules) 
or adding new pseudo-scalar and/or axial-vector fields to our Lagrangian can eliminate
the linear terms from the scattering amplitudes. For example, the relation
\begin{equation}\label{eq:SumRule-pipipipi}
   a_V^2 + \frac{3}{4}\alpha = 1
\end{equation}
removes the linear dependence on $s$ from the $\pi\pi\rightarrow\pi\pi$ amplitude.
The sum rule~(\ref{eq:SumRule-pipipipi}) links $a_V$ with $g''$ and $M_\rho$
via $\alpha=[2M_\rho/(vg'')]^2$. The resulting sets of $(a_V,g'')$ points
for given $M_\rho$'s are also shown in Fig.~\ref{fig:U-limits}. Obviously,
the vector resonance satisfying the existing experimental limits is able to 
follow the sum rule if its mass stays close or below $1.5\unit{TeV}$.

It is not possible to satisfy all sum rules resulting from the elastic and inelastic
$\pi\pi$ scatterings at the same time~\cite{CompositeHiggsSketch}. In addition, 
the sum rule motivated by the $\pi\pi\rightarrow h\rho_L$ amplitude reads
\begin{equation}\label{eq:SumRule-pipihrho}
   a_V=a_\rho.
\end{equation}
Unfortunately, this is the no-splitting condition not preferred by the data.
However, the unitarity-based conclusions reached in~\cite{CompositeHiggsSketch} rely on
the assumption~(\ref{eq:SumRule-pipihrho}). Thus, the results obtained in our analysis
call for the investigation of the inelastic unitarity limits when $a_V\neq a_\rho$. 
However, this is beyond the scope of this paper
and currently work in progress. \\


\section{The vector resonance mass limits and 
the upper bounds on \bm{$\sigma(pp\rightarrow \rho+X)\times\mbox{BR}$}}
\label{sec:LimitsVectorMassAndXS}

\begin{table*}[t]
\centering
\caption{\label{tab:ProdXSxBR1000}
         The production cross section times the branching ratio for different decay channels
         of the $1$, $1.5$, and $2\unit{TeV}$ vector resonances of our model 
         considering three different values of $g''$. The predictions for the $ZH$ and
         $WH$ channels are given for $a_V=1$ and $a_\rho=0$. No direct interactions of the
         vector resonance with fermions are assumed. The cross sections in the table
         are calculated for the $13\unit{TeV}$ pp collisions.
        } 
{\renewcommand{\arraystretch}{1.6}
\begin{tabular}{|c|c|c|c|c|}       
   \hline
   Channel & $g''$ & $\sigma(\rho^0)$ (pb) & Channel & $\sigma(\rho^+)+\sigma(\rho^-)$ (pb) \\
   \hline 
           & $10$ & $0.087$ $0.016$ $0.005$ &  & $0.156$ $0.030$ $0.009$ \\
   \cline{2-3}\cline{5-5}
      $WW$ & $15$ & $0.038$ $0.007$ $0.002$ & $ZW$ & $0.068$ $0.013$ $0.004$ \\
   \cline{2-3}\cline{5-5}
           & $20$ & $0.021$ $0.004$ $0.001$ & & $0.038$ $0.007$ $0.002$  \\
   \hline 
           & $10$ & ($71$ $2.8$ $0.2$)$\times 10^{-7}$ & & ($233$ $9.7$ $0.9$)$\times 10^{-7}$ \\
   \cline{2-3}\cline{5-5}
      $ZH$ & $15$ & ($31$ $1.2$ $0.1$)$\times 10^{-7}$ & $WH$ & ($103$ $4.3$ $0.4$)$\times 10^{-7}$ \\
   \cline{2-3}\cline{5-5}
           & $20$ & ($18$ $0.7$ $0.1$)$\times 10^{-7}$ & & ($\phantom{0}57$ $2.4$ $0.2$)$\times 10^{-7}$ \\
   \hline 
           & $10$ & ($427$ $16.3$ $1.5$)$\times 10^{-6}$ & & ($576$ $23.1$ $2.2$)$\times 10^{-6}$ \\
   \cline{2-3}\cline{5-5}
      $jj$ & $15$ & ($188$ $\phantom{0}7.2$ $0.7$)$\times 10^{-6}$ & $jj$ & ($253$ $10.2$ $1.0$)$\times 10^{-6}$ \\
   \cline{2-3}\cline{5-5}
          ($u,d,c,s,b$) & $20$ & ($105$ $\phantom{0}4.0$ $0.4$)$\times 10^{-6}$ & ($u,d,c,s$) & ($142$ $\phantom{0}5.7$ $0.5$)$\times 10^{-6}$ \\
   \hline 
           & $10$ & ($107$ $4.1$ $0.4$)$\times 10^{-6}$ & & ($191$ $7.8$ $0.7$)$\times 10^{-6}$ \\
   \cline{2-3}\cline{5-5}
      $e^+e^-+\mu^+\mu^-$ & $15$ & ($\phantom{0}47$ $1.8$ $0.2$)$\times 10^{-6}$ & $e\nu_e+\mu\nu_\mu$ & ($\phantom{0}84$ $3.4$ $0.3$)$\times 10^{-6}$ \\
   \cline{2-3}\cline{5-5}
           & $20$ & ($\phantom{0}26$ $1.0$ $0.1$)$\times 10^{-6}$ & & ($\phantom{0}47$ $1.9$ $0.2$)$\times 10^{-6}$ \\
   \hline 
           & $10$ & ($105$ $4.1$ $0.4$)$\times 10^{-6}$ & &  ($280$ $11.3$ $1.1$)$\times 10^{-6}$ \\
   \cline{2-3}\cline{5-5}
      $tt$ & $15$ & ($\phantom{0}46$ $1.8$ $0.2$)$\times 10^{-6}$ & $tb$ & ($123$ $\phantom{0}4.9$ $0.5$)$\times 10^{-6}$ \\
   \cline{2-3}\cline{5-5}
      & $20$ & ($\phantom{0}26$ $1.0$ $0.1$)$\times 10^{-6}$ & & ($\phantom{0}69$ $\phantom{0}2.8$ $0.3$)$\times 10^{-6}$ \\
   \hline 
\end{tabular}}
\end{table*}

Undeniably,  the search for new vector (and other) resonances has its rightful 
and important place in the ATLAS and CMS Collaboration's activities. While no discovery has 
been made, the direct exclusion limits constantly improve.  Unfortunately, the 
obtained limits are strongly model and parameter dependent. No wonder that the 
mass exclusion limits found in the literature cover only some of the interesting cases.
To the best of our knowledge, there are no exclusion mass limits applicable to the vector
resonance of the model studied in this paper.

One of the crucial factors on which the exclusion mass limit depends is the 
value of the vector resonance gauge coupling $g''$. As we saw in Fig.~\ref{fig:U-limits}, 
in our model, the superposition of the unitarity limits 
over the experimentally preferred region $0.77\leq a_V\leq 1.09$ suggests that 
considering $g''$ below about 10 is not well justified. We cannot rely on the 
predictions of our Lagrangian with experimentally allowed values of $a_V$ when 
$g''=\cO(1)$. This is not of great concern to us if the motivation for our model
stems from strongly interacting physics.
Independently of this restriction, if we narrowed our 
considerations to the tBESS interaction pattern of the vector resonance to 
fermions the low-energy limit from the EW precision measurements reads $g''\geq 
12$ at $95\%$~CL~\cite{tBESSepjc13}. This is an additional motivation for considering $g''$ 
values above 10. 

We can evaluate how the existing ATLAS and CMS data restrict our model when we compare 
the predictions of our model with the upper bounds
on the resonance production cross section times its branching ratio for various decay
channels. The bounds are rather model independent once
spin of the resonance under consideration is specified. Of course,
one should keep in mind that the calculations involved proceed under the assumption
of a narrow-width resonance.

In Table~\ref{tab:ProdXSxBR1000},
we present the cross section times branching ratios for various decay channels
of the considered model at the LHC collision energy of $13\unit{TeV}$.
The predictions are given for three different values of the resonance masses, namely
$1$, $1.5$, and $2\unit{TeV}$, and three values of $g''$, namely $10$, $15$, and $20$.
The $g''$ values were chosen to span the region allowed by the combination of the
limits considered and obtained in the previous sections. Note that for $g''>20$ our theory
not only runs into its perturbativity limit, but heavier vector resonances depart from 
the narrow width requirement as can be read off of the upper 
$x$-axes of Fig~\ref{fig:U-limits}.  

All model predictions quoted in Table~\ref{tab:ProdXSxBR1000}
correspond to the scenario when the decay of the vector resonance to fermions is
negligible.
In the tBESS-like fermion sector, this would correspond to turning off 
the direct coupling of the vector resonance to the third quark generation\footnote{
 In the tBESS model~\cite{tBESSepjc13}, there is no direct interaction of the vector resonance
 with the light fermions. The direct coupling of the vector resonance 
 to the third quark generation is parameterized by $b_L$ and $b_R$ and by the
 parameter $p$ which enables the splitting of the direct interactions of the right top
 and the right bottom quarks. Additional fermion sector parameters $\lambda_{L,R}$
 that have been introduced in the tBESS model are, for simplicity, kept at zero
 values throughout this paper.
}, i.e. $b_L=b_R=0$.
Consequently, in this scenario, all fermions can couple to
the vector resonance through the mixing-induced interactions only
and the decay widths of the neutral/charged vector resonances are dominated by
their decays to the EW gauge bosons; $\mathrm{BR}(WW/WZ)> 99\%$.

The experimental upper bounds on the cross section times branching ratio
in $WW$ channel read 0.033, 0.012, and $0.005\unit{pb}$
for $M_\rho=1$, 1.5, and $2\unit{TeV}$, respectively~\cite{ExpXS-AC062}.
This excludes the $1\unit{TeV}$ resonance for $g''\lessapprox  16$ and
$1.5\unit{TeV}$ resonance for $g''\lessapprox 12$. The $2\unit{TeV}$ resonance
is unrestricted for $g''\geq 10$.
Nevertheless, setting $b_{L,R}$ to their maximally low-energy precision data allowed values
--- $b_{L,R}\approx 0.1$, as found in~\cite{tBESSepjc13}\footnote{
While the preferred value of $p$ found in~\cite{tBESSepjc13} is about $0.25$,
its statistical preference over any other value of $p\in (0,1)$
is marginal. Therefore, we consider $p=1$ in our calculations of the tBESS-like fermion
sector contributions, i.e.\ the same strength of the vector resonance direct couplings
to the right top and bottom quarks.
} --- can lower
$\mathrm{BR}(WW)$ of the $1\unit{TeV}$ resonance down
to about $70\%$ for $g''=10$, to $30\%$ for $g''=15$,
and to $12\%$ for $g''=20$. In the $2\unit{TeV}$ resonance case, $\mathrm{BR}(WW)$ would be
lowered to about 97, 87, and $67\%$, respectively.
Thus, we can see that the fermionic interactions of the vector
resonance can noticeably decrease the predictions (and, thus, release the experimental
restrictions) of the model in this channel. The same effect can be expected to
occur in the $ZW$ channel.

Next, let us compare the predictions of our model to the experimental bounds in the $ZW$
channel. The most restrictive bounds in this channel read
$0.051\unit{pb}$~\cite{ExpXS-AC082}, $0.022\unit{pb}$~\cite{ExpXS-AC082}, 
and $0.009\unit{pb}$~\cite{ExpXS-AC055} for 
$M_\rho=1$, 1.5, and $2\unit{TeV}$, respectively. This excludes the $1\unit{TeV}$ 
resonance for $g''\lessapprox 17$ and $1.5\unit{TeV}$ resonance for $g''\lessapprox 12$.
Again, the $2\unit{TeV}$ resonance is unrestricted for $g''\geq 10$.
In the case of the tBESS-like fermion sector with $b_{L,R}=0.1$ and $p=1$, $\mathrm{BR}(WZ)$
gets lowered to about 71, 31, and $12\%$ for $g''=10$, $15$, and $20$, respectively,
when $M_\rho=1\unit{TeV}$. When $M_\rho=2\unit{TeV}$, the corresponding BR's read
97, 87, and $67\%$.

The exclusion limits from the $WW$/$WZ$ channels mentioned in the previous two paragraphs
were obtained from $13.2\unit{fb}^{-1}$~\cite{ExpXS-AC062,ExpXS-AC082} and 
$15.5\unit{fb}^{-1}$~\cite{ExpXS-AC055} of $13\unit{TeV}$ data.
For the sake of completeness, we should mention that the combined $WW+WZ$ analysis of
$3.2\unit{fb}^{-1}$ of $13\unit{TeV}$ data~\cite{ExpXS-AarXiv1} 
implies stronger limits $g''>13.6$ and $g''>10.6$
for $M_\rho=1.5$ and $2\unit{TeV}$, respectively.

At tree level, the $ZH$ and $WH$ decays of the vector resonances occur only when
$a_V\neq a_\rho$. In Table~\ref{tab:ProdXSxBR1000}, we present the predictions of
our model for $a_V=1$ and $a_\rho=0$. They can be used to derive the predictions
for preferred values of $a_V$ and $a_\rho$ that have been found in 
Section~\ref{subsec:CombinedAnalysis}
and quoted in Table~\ref{tab:numericalresults}. Following the formalism introduced in 
Section~\ref{sec:efflagr} we find that
\begin{equation}
   \Gamma_{ZH/WH}(a_V,a_\rho)=(a_\rho-a_V)^2\Gamma_{ZH/WH}(1,0).
\end{equation}
Then, taking into account the negligibility of the contributions of $\Gamma_{ZH}$
and $\Gamma_{WH}$ to the total decay widths of the neutral and charged resonances,
the $(a_\rho-a_V)^2$ scaling applies to the values of the production cross section times
the branching ratio for these channels. Thus, when $(a_V,a_\rho)=(0.93,-0.08)$,
the numbers in the $ZH/WH$ sector of Table~\ref{tab:ProdXSxBR1000} are to be
multiplied by $1.02$. When $(a_V,a_\rho)=(0.93,-1.68)$, $(-0.93,0.52)$, or
$(-0.93,2.11)$ the scaling factors are $6.81$, $2.10$, and $9.24$, respectively.
The experimental upper limits on the cross section times the branching ratio 
are $(0.070$~\cite{ExpXS-AarXiv2}, $0.032$~\cite{ExpXS-AarXiv2},
$0.013$~\cite{ExpXS-AC083}$)\unit{pb}$ for $M_\rho=(1.0,1.5,2.0)\unit{TeV}$, respectively, 
in the $ZH$ channel and 
$(0.12$~\cite{ExpXS-AarXiv2}, $0.035$~\cite{ExpXS-AC083}, $0.013$~\cite{ExpXS-AC083}$)\unit{pb}$ 
in the $WH$ channel.
The values predicted in these two channels by our model lie some four orders of magnitude
below the experimental upper limits. Thus, the upper limits provide no restriction 
within the considered range of $g''\in (10,20)$.

The same conclusions of no restrictions to our model can also be drawn for the remaining
decay channels from which the experimental upper bounds for the production cross section times
the branching ratio are available. Namely, the upper bounds for the charged $jj$ channel
read $0.210\unit{pb}$ and $0.088\unit{pb}$ for $M_\rho=1.5$ and $2\unit{TeV}$,
respectively, when $15.7\unit{fb}^{-1}$ of $13\unit{TeV}$ data~\cite{ExpXS-AC069} is 
processed\footnote{
The quoted upper bounds for all $jj$ channels include an acceptance factor.}. 
The combined neutral+charged $jj$ channel bounds based on $12.9\unit{fb}^{-1}$ of 
data~\cite{ExpXS-CP032}
are 1.20, 0.37, and $0.13\unit{pb}$ for $M_\rho = 1$, 1.5, and $2\unit{TeV}$, 
respectively. The bounds in the $e^+e^- + \mu^+\mu^-$
channel based on $13.3\unit{fb}^{-1}$ of data~\cite{ExpXS-AC045} 
read 1.30, 0.63, and $0.42\unit{fb}$ for $M_\rho = 1$, 1.5, and $2\unit{TeV}$, 
respectively. The bounds in the $e\nu_e + \mu\nu_\mu$
channel based on $13.3\unit{fb}^{-1}$ of data~\cite{ExpXS-AC061} 
read 4.8, 1.8, and $1.1\unit{fb}$ for $M_\rho = 1$, 1.5, and $2\unit{TeV}$, 
respectively. The bounds in the $tb$
channel based on $12.9\unit{fb}^{-1}$ of data~\cite{ExpXS-CP017} 
read 1.8, 0.55, and $0.23\unit{pb}$ for $M_\rho = 1$, 1.5, and $2\unit{TeV}$, 
respectively. Recall that the values in
Table~\ref{tab:ProdXSxBR1000} 
correspond to the scenario with no direct fermion interactions with the vector resonance.
In the tBESS-like fermion sector the $tb$ channel production would generally be higher.

Finally, there are also the upper bounds
for the $tt$ channel based on $3.2\unit{fb}^{-1}$ of data~\cite{ExpXS-AC014}: 
1.20, 0.33, and $0.17\unit{pb}$ for $M_\rho = 1$, 1.5, and $2\unit{TeV}$, respectively. 
When we compare these bounds with the predictions in Table~\ref{tab:ProdXSxBR1000}
we can see that for $g''\in (10,20)$ they do not restrict our model.
It applies even in the case of the tBESS-like fermion sector with $b_{L,R}=0.1$ and $p=1$.
Then the predictions in the $tt$ channel will rise to about $10^{-2}$,
$10^{-3}$, and $10^{-4}\unit{pb}$ for $M_\rho = 1$, 1.5, and $2\unit{TeV}$, respectively.


\section{Conclusions}
\label{sec:conclusions}

We have studied the experimental and unitarity limits on the parameters of the strong Higgs
sector of the phenomenological Lagrangian where beside the composite $125\unit{GeV}$
Higgs boson the $SU(2)_{L+R}$ triplet of composite vector resonances is 
explicitly present. The ESB sector of our effective Lagrangian has been based on the 
$SU(2)_L\times SU(2)_R\rightarrow SU(2)_{L+R}$ non-linear 
sigma model while the scalar resonance has been introduced 
as the $SU(2)_{L+R}$ singlet. The vector resonance has been
built in employing the hidden local symmetry approach. 

For the interactions of the Higgs boson with the EW gauge fields, the vector triplet,
and the top quark the assumed symmetries allow one to introduce 
three free independent modification factors, $a_V$, $a_\rho$, and $c_t$.
After the transformation from the flavor to mass gauge-boson eigenstate basis, the first
two factors combine into modifiers $c_Z$, $c_W$, $c_{\rho^0}$, $c_{\rho^\pm}$,
$c_{Z\rho^0}$, and $c_{W\rho^\pm}$ of the vertices $hZZ$, $hW^+W^-$, $h\rho^0\rho^0$,
$h\rho^\pm\rho^\pm$, $hZ\rho^0$, and $hW^\pm\rho^\mp$, respectively.
If $r=a_\rho/a_V\neq 1$, then $c_Z$ differs from $c_W$. Nevertheless, for quite a large
interval of $r$'s around one, e.g., $|r|\leq 3$, the effect is very small. 
The corrections to the custodial symmetry protected rho parameter induced
by the differing $a_V$ and $a_\rho$ are negligible and well within the experimental limits.
For many phenomenological considerations,
the approximations $c_Z=c_W=a_V$, $c_{\rho^0}=c_{\rho^\pm} = a_\rho$, and
$c_{Z\rho^0}=c_{W\rho^\pm} = 0$ are satisfactory over quite a large region of $r$ values
and for all relevant values of $g''$ and $M_\rho$.

The limits on the free Higgs coupling factors $a_V$, $a_\rho$, and $c_t$ have been calculated
using constraints on the kappa parameters of the interim framework. The constraints were 
obtained in the recent ATLAS+CMS Collaborations analysis of various Higgs-related processes
based on data from 2011 and 2012.
We have used the fitting scenario where no non-SM decays of the Higgs is assumed
and where the branching ratio of invisible and/or undetected decay products is zero.
In addition, new particles in loops are allowed. Out of seven free 
parameters in this scenario --- loop-level $\kappa_g$, $\kappa_\gamma$, and
tree-level $\kappa_W$, $\kappa_Z$, $\kappa_t$, $\kappa_b$, $\kappa_\tau$ --- we have used all 
but the last two to find the restrictions on $a_V$, $a_\rho$, and $c_t$;
$\kappa_b$ and $\kappa_\tau$ have been ignored because their impact
on the fitting parameters was negligible.

By fitting the kappas we have established that the simple case of $a_\rho=a_V$ is strongly
disfavored by
the data. There are four triplets of the best-fit values of $a_V$, $a_\rho$, and $c_t$ that
can satisfy the fitted data. Namely, $(a_V,a_\rho,c_t) = (0.93,-0.08,0.85),\;$ 
$(0.93,-1.68,0.85)$, $(-0.93,0.52,0.85)$, 
and $(-0.93,2.11,0.85)$. The 1-sigma deviation (i.e., $\chi^2_\mathrm{min}+1$)
for $a_V$ at all four best values is the same: $\pm 0.08$. It also applies to $c_t$. Its
1-sigma errors read $^{+0.11}_{-0.12}$ for each best value of $c_t$.
As far as $a_\rho$ is concerned its 1-sigma deviations
differ slightly at each of the best-fit values. However, overall they do not exceed $\pm 0.17$.
The hypothesis backing for all four fits
is the same and amounts to $12\%$. The tied score might be tilted in favor of one of the fits
once $\kappa_{Z\gamma}$ gets measured more precisely. 

Using scattering amplitudes of the longitudinal EW gauge bosons in the Equivalence
theorem approximation we have studied the restrictions of the usability of our
phenomenological Lagrangian imposed by the unitarity limits when the data preferred 
Higgs couplings obtained in this paper are considered. We have found from the 
$\pi\pi\rightarrow\pi\pi$ scattering that for
$M_\rho=1\unit{TeV}$ unitarity holds up to at least $\Lambda=3\unit{TeV}$ 
when $0.77\leq a_V\leq 1.09$ and $12\leq g''\leq 25$. As $M_\rho$ grows the region 
where unitarity holds shrinks toward higher $g''$ and lower $a_V$. Even if $M_\rho=2\unit{TeV}$,
the considered model is well below the unitarity limit at significant portions of
the experimentally allowed region of $a_V$. Nevertheless, these conclusions should be
complemented by a similar analysis of the $\pi\pi$ scattering with $h$ and $\rho_L$
in the final state where $a_V$ and $a_\rho$ would be independent parameters, thus allowing also
for $a_V\neq a_\rho$ preferred by the data.

Our calculations show that the masses
in the range $1\unit{TeV}\leq M_\rho\leq 2\unit{TeV}$ are not excluded
in parts or even full parameter space of our theory. When the model's
predictions face the upper bounds on the production cross section
times branching ratio in different decay channels the $1\unit{TeV}$ resonance
gets excluded in the $WW$ channel when $g''\lessapprox 16$ and in the $WZ$ channel
when $g''\lessapprox 17$. Both, the $WW$ and $WZ$ channel measurements exclude
the $1.5\unit{TeV}$ resonance when $g''\lessapprox 14$. These restriction can
get weakened once the direct interactions of the vector resonance with the fermion
sector are introduced. None of the other reviewed decay channels excludes
our model, at least when $10\leq g''\leq 20$.

In the view of the results obtained in this paper we would conclude that even
such a simplistic effective description of possible early phenomenology of strong BSM physics
as the one studied here is capable to accommodate the existing data.

\begin{acknowledgements}
We would like to thank F.~Riva for useful discussions. The work was
supported by the Grants LM2015058 and LG15052 of the Ministry of Education, 
Youth and Sports of the Czech Republic.
J.J.\ was also supported by the NSP grant of the Slovak Republic. 
M.G.\ was supported by the Slovak CERN Fund. We would also like to thank the
Slovak Institute for Basic Research for their support.
\end{acknowledgements}

\appendix

\section{Some definitions}
\label{app:definitions}

In this Appendix, we show definitions of some of the quantities used in Section~\ref{sec:efflagr}
to express the parts of our phenomenological Lagrangian relevant to this paper.
All details regarding the Lagrangian structure and how it was built can be 
found in~\cite{tBESSepjc13,tBESSprd11}.

The field strength tensors of the $SU(2)_L\times U(1)_Y\times SU(2)_\mathrm{HLS}$ gauge fields 
are defined as
\begin{eqnarray}
  \BW_{\mu\nu} &=& \pard_\mu \BW_\nu - \pard_\nu \BW_\mu + \Comm{\BW_\mu}{\BW_\nu},
  \label{Wmunu}\\
  \BB_{\mu\nu} &=& \pard_\mu \BB_\nu - \pard_\nu \BB_\mu,
  \label{Bmunu}\\
  \BV_{\mu\nu} &=& \pard_\mu \BV_\nu - \pard_\nu \BV_\mu + \Comm{\BV_\mu}{\BV_\nu},
  \label{Vmunu}
\end{eqnarray}
where $\BW_\mu=i g W_\mu^a\tau^a$,
$\BB_\mu=i g' B_\mu\tau^3$, and $\BV_\mu=i\frac{g''}{2}V_\mu^a\tau^a$
with the gauge couplings $g,g',\mathrm{\ and\ }g''$, respectively.

The ESB sector contains six unphysical real scalar fields, would-be Goldstone
bosons of the model's spontaneous symmetry breaking.
The six real scalar fields
$\vphi_L^a(x), \vphi_R^a(x),\; a=1,2,3$, are introduced as
parameters of the~$SU(2)_L\times SU(2)_R$ group elements in
the exp-form
$\xi(\vec{\vphi}_{L,R})=\exp(i\vec{\vphi}_{L,R}\vec{\tau}/v)\in SU(2)_{L,R}$
where $\vec{\vphi}=(\vphi^1,\vphi^2,\vphi^3)$.
The quantities $\bar{\omega}_\mu^{\parallel}$ and $\bar{\omega}_\mu^{\perp}$
are, respectively, 
the $SU(2)_{L-R}$ and $SU(2)_{L+R}$ projections of the gauged Maurer--Cartan 1-form,
\begin{eqnarray}
   \bar{\omega}_\mu^{\parallel} &=& \omega_\mu^{\parallel}+
   \frac{1}{2}\left(\xi_L^\dagger\BW_\mu\xi_L+\xi_R^\dagger\BB_\mu\xi_R\right)-
   \BV_\mu,
   \label{eq:gaugeMCparallel}\\
   \bar{\omega}_\mu^{\perp} &=& \omega_\mu^{\perp}+
   \frac{1}{2}\left(\xi_L^\dagger\BW_\mu\xi_L-\xi_R^\dagger\BB_\mu\xi_R\right),
   \label{eq:gaugeMCperp}
\end{eqnarray}
where $\omega_\mu^{\parallel,\perp}=
(\xi_L^\dagger\pard_\mu\xi_L\pm\xi_R^\dagger\pard_\mu\xi_R)/2$.





\begin{thebibliography}{99}

\bibitem{125GeVBosonDiscoveryATLAS}
G.~Aad \textit{et al.} (ATLAS Collaboration), Phys. Lett. B
\textbf{716}, 1 (2012).

\bibitem{125GeVBosonDiscoveryCMS}
S.~Chatrchyan \textit{et al.} (CMS Collaboration), Phys. Lett. B
\textbf{716}, 30 (2012).

\bibitem{Barbieri_etal07}
R. Barbieri, B. Bellazini, V. S. Rychkov and A. Varagnolo, Phys. Rev. \textbf{D76} (2007) 115008.

\bibitem{CsakiFalkowskiWeiler08}
C. Csaki, A. Falkowski and A. Weiler, JHEP 0809 (2008) 008.

\bibitem{Contino10}
R. Contino, arXiv:hep-ph/1005.4269.

\bibitem{Contino_etal11}
R. Contino, D. Marzocca, D. Pappadopulo and R. Rattazzi, JHEP 1110 (2011) 081.

\bibitem{PomarolRiva12}
A. Pomarol and F. Riva, JHEP 1208 (2012) 135.

\bibitem{PappadopuloThammTorre13}
D. Pappadopulo, A. Thamm and R. Torre, JHEP 1307 (2013) 058.

\bibitem{Montull_etall13}
M. Montull, F. Riva, E. Salvioni and R. Torre, Phys. Rev. \textbf{D88}, 095006 (2013).

\bibitem{PanicoWulzer16}
G. Panico and A. Wulzer, Lect. Notes Phys. 913 (2016) 1, arXiv:1506.01961.

\bibitem{BardeenHillLindner90}
W. A. Bardeen, C. T. Hill and M. Lindner, Phys. Rev. \textbf{D41}, 1647 (1990).

\bibitem{Giudice_etal07}
G. F. Giudice, C. Grojean, A. Pomarol and R. Rattazzi, JHEP 0706 (2007) 045.

\bibitem{Foadi_etal07}
R. Foadi, M. T. Frandsen, T. A. Ryttov and F. Sannino, Phys. Rev. D 76 (2007) 055005.

\bibitem{RyttovSannino08}
T. A. Ryttov and F. Sannino, Phys. Rev. D 78 (2008) 115010.

\bibitem{Sannino09}
F. Sannino, Acta Phys. Polon. B 40 (2009) 3533.

\bibitem{Zerwekh10}
A. R. Zerwekh, Mod. Phys. Lett. A 25 (2010) 423.

\bibitem{HernandezTorre10}
A. E. C\'{a}rcamo Hern\'{a}ndez and R. Torre, Nucl. Phys. B 841 (2010) 188.

\bibitem{BurdmanHaluch11}
G. Burdman and C. E. F. Haluch, JHEP 1112 (2011) 038.

\bibitem{HapolaSannino11}
T. Hapola and F. Sannino, Mod. Phys. Lett. A 26 (2011) 2313.

\bibitem{Hernandez_etal12}
A. E. C\'{a}rcamo Hern\'{a}ndez, C. O. Dib, N. Neill H and A. R. Zerwekh, JHEP 1202 (2012) 132.

\bibitem{Foadi_etal13}
R. Foadi, M. T. Frandsen and F. Sannino, Phys. Rev. D 87 (2013) 095001.

\bibitem{Contino_etal13}
R. Contino, M. Ghezzi, C. Grojean, M. Muhlleitner and M. Spira, JHEP 1307 (2013) 035.

\bibitem{Castillo-Felisola_etal13}
O. Castillo-Felisola, C. Corral, M. Gonz\'{a}lez, G. Moreno, N. A. Neill, 
F. Rojas, J. Zamora and A. R. Zerwekh, Eur. Phys. J. C 73 (2013) 2669.

\bibitem{HernandezDibZerwekh14}
A. E. C\'{a}rcamo Hern\'{a}ndez, C. O. Dib and A. R. Zerwekh, Eur. Phys. J. C 74, 2822 (2014);
Nucl. Part. Phys. Proc. 267-269 (2015) 35.

\bibitem{Pappadopulo_etal14}
D. Pappadopulo, A. Thamm, R. Torre and A. Wulzer, JHEP 1409 (2014) 060.

\bibitem{Belyaev_etal14}
A. Belyaev, M. S. Brown, R. Foadi and M. T. Frandsen, Phys.Rev. D 90 (2014) 035012.


\bibitem{LeeQuiggThacker77}
B.~W.~Lee, C.~Quigg, and H.~B.~Thacker, Phys. Rev. D \textbf{16},
1519 (1977); 
Phys. Rev. Lett. \textbf{38} (1977) 883.

\bibitem{Chanowitz85}
M.~S.~Chanowitz and M.~K. Gaillard, Nucl. Phys. \textbf{B261}, 379
(1985).

\bibitem{CompositeHiggsSketch}
B.~Bellazzini et al, JHEP
11 (2012) 003.

\bibitem{HLS}
M.~Bando, T.~Kugo, and K.~Yamawaki, Phys. Rep. \textbf{164}, 217
(1988).

\bibitem{3siteHiggslessModel}
R.S.~Chivukula et al, Phys. Rev. D
\textbf{74}, 075011 (2006).

%
%
%
%

\bibitem{tBESSprd11}
M.~Gintner, J.~Jur\'{a}\v{n}, and I.~Melo, Phys. Rev. D
\textbf{84}, 035013 (2011).

\bibitem{tBESSepjc13}
M.~Gintner, J.~Jur\'{a}\v{n}, Eur. Phys. J. C
\textbf{73}, 2577 (2013).

\bibitem{BESS}
R.~Casalbuoni, S.~De~Curtis, D.~Dominici, and R.~Gatto, Phys.
Lett. \textbf{155B}, 95 (1985); Nucl. Phys. \textbf{B282}, 235
(1987); R.~Casalbuoni, P.~Chiappetta, S.~De~Curtis, F.~Feruglio,
R.~Gatto, B.~Mele, and J.~Terron, Phys. Lett. \textbf{B249}, 130
(1990).

\bibitem{ATLAS-CMS-HiggsDataNew}
The ATLAS and CMS Collaborations, arXiv:1606.02266.

\bibitem{kappa-framework}
The LHC Higgs Cross Section Working Group, arXiv:1307.1347.

\bibitem{HandbookLHCHiggsXS1}
The LHC Higgs Cross Section Working Group, arXiv:1101.0593.

\bibitem{ATLAS-HiggsData}
ATLAS Collaboration, Eur. Phys. J. C \textbf{76}, 6 (2016).

%
%
%
%
%

\bibitem{ExpXS-AC062}
ATLAS Collaboration, ATLAS-CONF-2016-062.

\bibitem{ExpXS-AC082}
ATLAS Collaboration, ATLAS-CONF-2016-082.

\bibitem{ExpXS-AC055}
ATLAS Collaboration, ATLAS-CONF-2016-055.

\bibitem{ExpXS-AarXiv1}
ATLAS Collaboration, arXiv:1606.04833.

\bibitem{ExpXS-AarXiv2}
ATLAS Collaboration, arXiv:1607.05621.

\bibitem{ExpXS-AC083}
ATLAS Collaboration, ATLAS-CONF-2016-083.

\bibitem{ExpXS-AC069}
ATLAS Collaboration, ATLAS-CONF-2016-069.

\bibitem{ExpXS-CP032}
CMS Collaboration, CMS-PAS-EXO-16-032.

\bibitem{ExpXS-AC045}
ATLAS Collaboration, ATLAS-CONF-2016-045.

\bibitem{ExpXS-AC061}
ATLAS Collaboration, ATLAS-CONF-2016-061.

\bibitem{ExpXS-CP017}
CMS Collaboration, CMS-PAS-B2G-16-017.

\bibitem{ExpXS-AC014}
ATLAS Collaboration, ATLAS-CONF-2016-014.




\end{thebibliography}
\end{document}